\documentclass[a4paper,11pt]{article}
\pdfoutput=1 

\usepackage{amsthm,booktabs,integrable,jheparxiv,xparse,stmaryrd}

\title{The Celestial Chiral Algebra of Self-Dual Gravity on Eguchi-Hanson Space}

\author[a]{Roland Bittleston,}
\author[b]{Simon Heuveline}
\author[b]{and David Skinner}

\affiliation[a]{Perimeter Institute for Theoretical Physics,\\ 51 Caroline Street, Waterloo, Ontario, Canada\vspace{0.1cm}}
\emailAdd{rbittleston@perimeterinstitute.ca}

\affiliation[b]{Dept. Applied Maths \& Theoretical Physics,\\ University of Cambridge, Wilberforce Road, CB3 0WA, United Kingdom\vspace{0.1cm}}
\emailAdd{sph48@cam.ac.uk}
\emailAdd{d.b.skinner@damtp.cam.ac.uk}

\begin{document}

\abstract{
We consider the twistor description of classical self-dual Einstein gravity in the presence of a defect operator wrapping a certain $\CP^1$. The backreaction of this defect deforms the flat twistor space to that of Eguchi-Hanson space. We show that the celestial chiral algebra of self-dual gravity on the Eguchi-Hanson background is likewise deformed to become the loop algebra of a certain scaling limit of the family of $W(\mu)$-algebras, where the scaling limit is controlled by the radius of the Eguchi-Hanson core. We construct this algebra by computing the Poisson algebra of holomorphic functions on the deformed twistor space, and check this result with a space-time calculation of the leading contribution to the gravitational splitting function. The loop algebra of a general $W(\mu)$-algebra (away from the scaling limit) similarly arises as the celestial chiral algebra of Moyal-deformed self-dual gravity on Eguchi-Hanson space. We also obtain corresponding results for self-dual Yang-Mills.
}

\maketitle

\flushbottom


\section{Introduction}

Celestial holography posits that a gravitational theory in a four dimensional asymptotically flat space-time should be holographically dual to an exotic two dimensional CFT living on a Riemann sphere. At present, the most concrete example of this is due to Costello, Paquette \& Sharma~\cite{Costello:2022jpg,Costello:2023hmi} who combine ideas from twisted holography~\cite{Costello:2018zrm,Costello:2020jbh} with Penrose's twistor theory~\cite{Penrose:1968me,Penrose:1976js} to show that scalar-flat K{\"a}hler gravity, coupled to a 4-dimensional WZW model with $G=\gSO(8)$, in an asymptotically flat space known as Burns space~\cite{Burns:1986,LeBrun:1991} is holographically dual to a 2d CFT that arises as a chiral twist~\cite{Beem:2013sza} of a 4d $\cN=2$ gauge theory.

\medskip

\begin{figure}[t!]
	\centering
 	\includegraphics[scale=0.33]{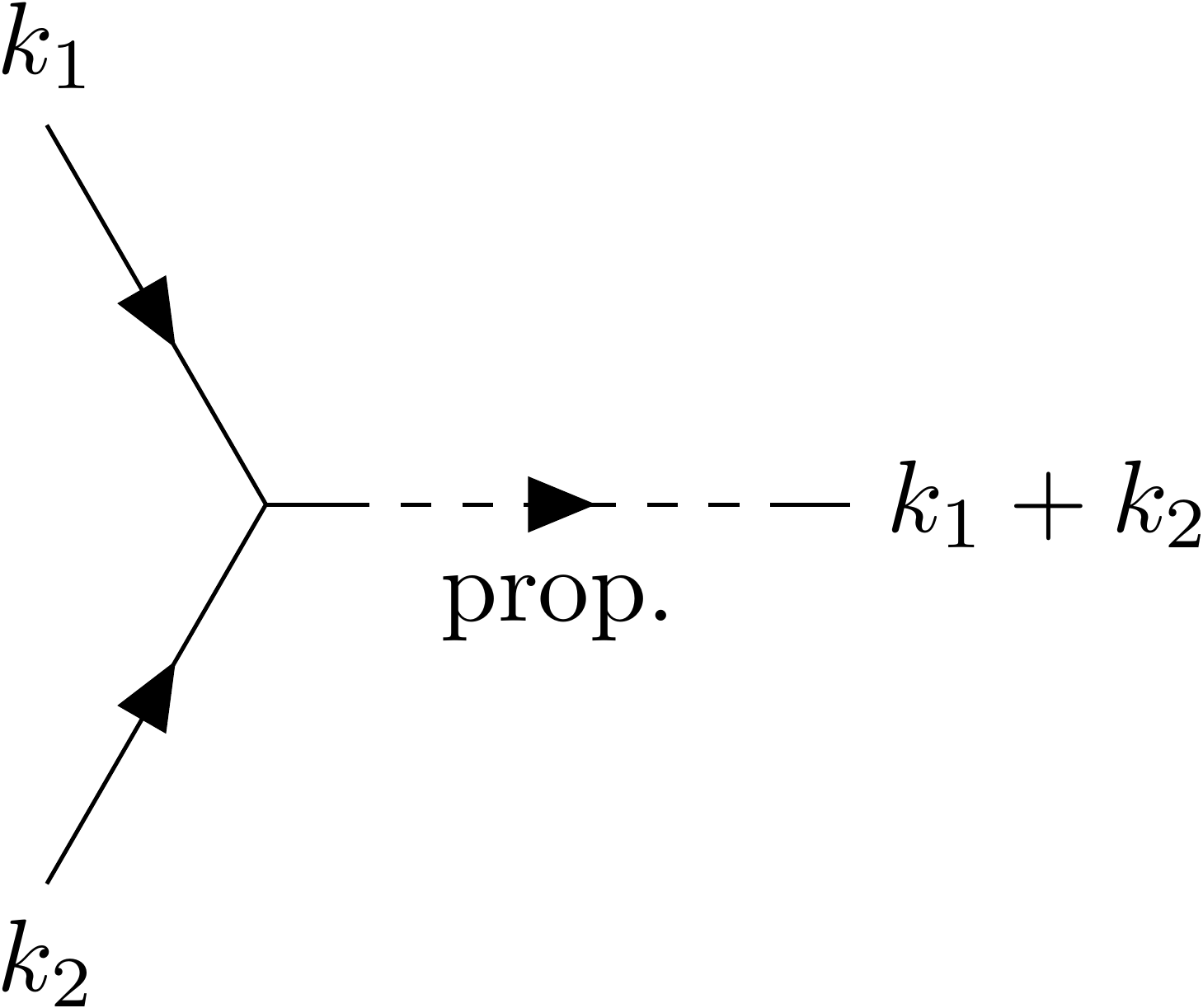}
	\caption{\emph{Tree diagram responsible for the singularity in a graviton amplitude as the momenta $k_1,k_2$ of two positive helicity external states become collinear. A simple pole is generated when the when the propagator goes on-shell.}} \label{fig:splitting}
\end{figure}

Aside from studying this specific example, one can try to deduce generic properties of celestial duals by studying universal features of gravitational scattering amplitudes. An important property here are the singularities such amplitudes develop as the momenta of two external massless states become collinear. As shown long ago \cite{Weinberg:1965nx, Kulish:1970ut,Altarelli:1977zs,Berends:1987me,Mangano:1990by}, these are governed by \emph{splitting functions} which in perturbation theory arise from the diagram in figure~\ref{fig:splitting}. They signal a breakdown in the Fock space description of the Hilbert space of the massless field (see {\it e.g.}~\cite{Prabhu:2022zcr} for a recent discussion) and, in the context of celestial holography, shed light on the structure of the OPE between operators in the 2d CFT representing modes of bulk fields such as the graviton.

The form of the splitting functions is well known. In particular, for gravity in flat space-time one has
\be \label{eq:gravity-flat-split}
    \text{Split}_+^{\rm tree} = - \frac{[12]}{\la 12\ra} 
\ee
for the tree-level splitting function between a pair of positive helicity gravitons of momenta $k_1 = |1\ra[1|$ and $k_2 = |2\ra[2|$ \cite{Bern:1998xc,Bern:1998sv}. In the holomorphic collinear limit $\la12\ra\to0$, the splitting function describes the merging of two on-shell gravitons into a new one,  thus turning the space of on-shell gravitons (linearized fluctuations of the metric) into an algebra. The conformally soft modes of a positive helicity graviton are defined~\cite{Kapec:2016jld, Pasterski:2017kqt,Pasterski:2017ylz} as  residues of the Mellin transform of its wavefunction $\delta\Theta_k(x) = \exp(\im x\cdot k)/\la\alpha\kappa\ra^4$ by
\be \label{eq:conformally-soft-def}
\mathrm{Res}_{\Delta=-m}\left[\int_0^\infty\dif\omega\,\omega^{\Delta-1}\,\delta\Theta_k(x)\right] = \im^m\sum_{p+q=m}\frac{\tilde z^q}{p!q!}\,w[p,q](z)\,,
\ee
for  $m\in\bbN_0$, where we have parametrised its momentum spinors as $|\kappa\ra = \sqrt{\omega}(1,z)$, $|\kappa] = \sqrt{\omega}(1,\tilde z)$. The splitting function~\eqref{eq:gravity-flat-split} then leads to the OPE
\be
\label{eq:ham-C2-def}
w[p,q](z)\,w[r,s](0)\sim-\frac{ps-qr}{2z}\,  w[p\!+\!r\!-\!1,q\!+\!s\!-\!1](0)
\ee
between these conformally soft modes. This algebra was shown in~\cite{Guevara:2021abz,Strominger:2021lvk,Ball:2021tmb} to be a perturbatively exact symmetry of the scattering  amplitudes of self-dual gravity. If gravitational form factors are considered in the self-dual theory, or if one goes beyond self-dual gravity to full Einstein gravity, the algebra receives loop-level corrections studied in~\cite{Bittleston:2022jeq}. In the context of self-dual gravity, these corrections are best understood in the presence of an additional axion field whose role is to cancel diffeomorphism anomalies in twistor space~\cite{Costello:2022upu, Bittleston:2022nfr}. 

It is important to note that the indices $p,q,\ldots$ in~\eqref{eq:ham-C2-def} run over all integers $\geq0$. Since $w[p,q](z)$ transforms in the  $(p+q+1)$-dimensional representation of the $\fsl_2$ generated by the zero modes of $w[2,0]$, $w[1,1]$ and $w[0,2]$, the conformally soft modes encompass all (integer and half-integer) spins. The algebra~\eqref{eq:ham-C2-def} can be identified with $\cL\ham(\bbC^2)$; the loop algebra of the Lie algebra of holomorphic Hamiltonian vector fields on $\bbC^2$, equipped with the Poisson bracket $\{f,g\}=\p_uf\p_vg - \p_vf\p_ug$. The relevance of this algebra to self-dual gravity was recognised long ago~\cite{Penrose:1976js,Penrose:1972ia} and forms the basis of the non-linear graviton construction of twistor theory. The fact that the occurrence of $\cL\ham(\bbC^2)$ in twistor theory should be identified with its occurrence in scattering amplitudes was explained in~\cite{Adamo:2021lrv} by considering graviton vertex operators in the twistor sigma model~\cite{Adamo:2021bej}. As usual, these vertex operators represent infinitesimal deformations of the background (twistor) space, generating a non-linear graviton, while their OPEs in the sigma model immediately yield~\eqref{eq:ham-C2-def}.

\medskip

We emphasise that the celestial chiral algebra $\cL\ham(\bbC^2)$ is \emph{not} the loop algebra of the wedge subalgebra of $w_{1+\infty}$, denoted $\cL w_\wedge$, but rather it contains $\cL w_\wedge$ as a subalgebra. Indeed, while $\cL w_\wedge$ is defined by exactly the same relations as~\eqref{eq:ham-C2-def}, it contains only the generators $w[p,q]$ with $p+q\equiv0\Mod{2}$. As is well known (see, {\it e.g.},~\cite{Dixmier:1973quo,Feigin:1988lie,Pope:1989sr,Bergshoeff:1989ns,Bordemann:1989zi,Pope:1991ig}), the wedge subalgebra $w_\wedge\subset w_{1+\infty}$ admits a family $W(\mu)$ of deformations, defined by
\bea \label{eq:W(mu)-commutator}
&[\widetilde W[p,q],\widetilde W[r,s]] \\
&= \sum_{\ell\geq0}\fq^{2\ell}R_{2\ell+1}(p,q,r,s)\Psi_{2\ell+1}\bigg(\frac{p\!+\!q}{2},\frac{r\!+\!s}{2};\sigma\bigg)\widetilde W[p\!+\!r-\!2\ell\!-\!1,q\!+\!s\!-\!2\ell\!-\!1]\,,
\eea
where
\be R_\ell(p,q,r,s) = \frac{1}{\ell!}\sum_{k=0}^\ell(-)^k\binom{\ell}{k}[p]_{\ell-k}[q]_k[r]_k[s]_{\ell-k}\,, \ee
$\Psi$ is the hypergeometric function
\be 
\label{eq:Psi-ell} 
\Psi_\ell(m,n;\sigma) = 
\pFq{4}{3}{-1/2\!-\!2\sigma,3/2\!+\!2\sigma,-\!\ell/2,(1\!-\!\ell)/2}{1/2\!-\!m,1/2\!-\!n,m\!+\!n\!+\!3/2\!-\!\ell}{1}
\ee
and $\sigma$ is a real parameter. Since the hypergeometric function is unchanged by the replacement $\sigma\mapsto -1/2-\sigma$, the algebras are properly labelled by $\mu=\sigma(\sigma+1)$. The formal parameter $\fq$ controls the deformation, with $\lim_{\fq\to0} W(\mu) = w_\wedge$; the algebras with different $\fq\neq0$ are all isomorphic. 

One reason these $W$-algebras are of potential interest to celestial holography is that, while $w_{1+\infty}$ is known to arise as the {\it classical} algebra of various 2d CFTs, at the quantum level this is usually deformed to a $W$-algebra, with $\fq$ playing the role of $\hbar$~\cite{Bakas:1989xu,Pope:1989ew,Bergshoeff:1991un}. However, the CCA of gravitational scattering amplitudes involves not $w_\wedge$, but $\ham(\bbC^2)$. For generic values of $\sigma$, the hypergeometric functions~\eqref{eq:Psi-ell} with $\ell\geq1$ have poles at half-integer values of $m,n$, so that a generic $W(\mu)$ cannot be lifted to a deformation of $\ham(\bbC^2)$. The only exception -- the unique deformation\footnote{At least, within the category of Lie algebras \cite{Etingof:2023new}. As emphasised to us by Strominger, it's also possible that further deformations of the \emph{loop} algebra $\cL\ham(\bbC^2)$ do exist.}  of $\ham(\bbC^2)$ --  occurs at $\mu=-3/16$ ($\sigma=-1/4,-3/4$), where $\Psi_\ell(m,n;-1/4)$ has no poles even for half-integer $m,n$. $W(-3/16)$ may be obtained by replacing the Poisson bracket with its corresponding Moyal bracket and is known as the \emph{symplecton algebra} (also known as the Weyl algebra). Indeed, in~\cite{Bu:2022iak} it was shown that this symplecton algebra arises from collinear singularities of graviton scattering amplitudes in a non-commutative self-dual gravity. 

\medskip

It is natural to ask whether the generic $W(\mu)$-algebras could still have a role in celestial holography. In this paper we show that they do, but for celestial holography on the background of Eguchi-Hanson space rather than flat space-time. As the reader is no doubt aware, Eguchi-Hanson space~\cite{Eguchi:1978xp} can be viewed as $T^*\CP^1$ endowed with a Ricci flat metric. Asymptotically, the metric is locally that of flat space, but globally as $|x|\to\infty$ Eguchi-Hanson space approaches the orbifold $\bbR^4/\bbZ_2$. Indeed, the whole space is naturally a deformation of this orbifold, with the deformation parameter governing the size of the $\CP^1$ base. 

To summarise the results of this paper, the significance of these facts for celestial holography is this. Only the modes $w[p,q](z)$ with $p+q\equiv0\Mod{2}$ survive on the orbifold. Thus, taking the $\bbZ_2$ quotient reduces the celestial chiral algebra of self-dual gravity from $\cL\ham(\bbC^2)$ to its subalgebra $\cL w_\wedge$. Deforming from the orbifold to Eguchi-Hanson space likewise deforms $\cL w_\wedge$ to an algebra we call $\cL W(\infty)$. This is the loop algebra of a scaling limit of the $W(\mu)$-family, with $\fq\to0$ and $\mu\to\infty$, but with $\fq^2\mu$ held fixed in terms of the radius of the Eguchi-Hanson space.\footnote{Our $W(\infty)$ is not to be confused with the wedge subalgebra of $W_\infty$.} Alternatively, it's also the complexification of $\mathfrak{sdiff}(S^2)$, the Lie algebra of area preserving vector fields on the sphere. Finally, the loop algebra of the wedge subalgebra of a generic member of the $W(\mu)$ family arises as the celestial chiral algebra of non-commutative self-dual gravity on Eguchi-Hanson space. The fact that, without turning on non-commutativity, $\ham(\bbC^2)$ has no non-trivial deformations can thus be seen as an algebraic counterpart of the Gibbons-Pope theorem~\cite{Gibbons:1979xn} that the only Ricci flat, asymptotically flat space-time is flat space itself.  It is clear that many of the considerations of this paper extend to the $ADE$ series of ALE spaces. These will be discussed in the forthcoming paper~\cite{Bittleston:2023soon}.

\medskip

This paper is structured as follows. We begin in section~\ref{sec:sdgravity-twistors} with a brief review of the (classical) twistor action for self-dual gravity, initially focusing on perturbation theory around the twistor space $\PT$ of flat $\bbR^4$. In section~\ref{sec:deforming-twistor-space}, as in twisted holography~\cite{Costello:2018zrm} and the top-down celestial holographic model of~\cite{Costello:2022jpg}, we couple this theory to a defect wrapping a complex curve $\CP^1\subset\PT$. We show that the backreaction sourced by this defect deforms $\PT$ to the twistor space $\cPT$ of Eguchi-Hanson space. In an effort to keep this paper self-contained, we review the important features of $\cPT$, including its projection $\CP^1$, the weight 2 symplectic structure on the fibres of this projection, and the 4-parameter family of holomorphic sections, from which the space-time and Eguchi-Hanson metric may be recovered. In section~\ref{sec:twistor-algebra} we use the twistor space to construct (at the classical level) the celestial chiral algebra (CCA) of self-dual gravity on Eguchi-Hanson space, identifying this with the loop algebra of the wedge subalgebra of a scaling limit of $W(\mu)$, as discussed above. We also consider the CCA of self-dual Yang-Mills for a complex semisimple Lie algebra $\fg$. On flat space it's CCA is $\cL\fg[\bbC^2]$, the loop algebra of the Lie algebra of polynomial maps from $\bbC^2$ into $\fg$. We show that on Eguchi-Hanson this is deformed to an algebra we denote $S_\wedge(\infty)$, the detailed structure of which is given in section \ref{subsec:twistor-CCA-SDYM}.

The results of section~\ref{sec:twistor-algebra} essentially follow just from the ring of holomorphic functions on the fibres of $\cPT\to\CP^1$. In section~\ref{sec:space-time-algebra} we recover the same algebras by calculating, entirely on space-time, the gravitational and Yang-Mills splitting functions for scattering on the Eguchi-Hanson background. In section~\ref{sec:non-commutativity}, we allow the twistor space to become non-commutative, showing that the CCA now corresponds to the loop algebra of the wedge subalgebra of a generic $W(\mu)$ algebra, with $\fq$ governing the non-commutativity and $\fq^2\mu$ determined by the Eguchi-Hanson radius. We conclude in section~\ref{sec:discussion} with a brief discussion of some open directions. 

\medskip

In the past a number of connections have been made between self-dual gravity and $W$-algebras. In particular, the works \cite{Park:1989vq,Park:1989fz,Ooguri:1991fp} connect self-dual gravity and loop algebras of $w_\infty/W_\infty$ (\emph{not} their wedge subalgebras) using the results of \cite{Bakas:1989xu}. We believe the appearance of full $W$-algebras, rather than their wedge subalgebras, is because the authors consider local rather than global holomorphic Hamiltonian vector fields on the twistor fibres. On the other hand, in \cite{Dunajski:2000iq} the authors compute the hidden symmetry algebra of the 2\textsuperscript{nd} Pleba\'{n}ski equation, obtaining the Lie algebra of loops into the split extension of a Lie algebra of Hamiltonian vector fields by a certain module. We expect that restricting to the unextended algebra and specialising to the case of Eguchi-Hanson would recover the $W(\infty)$ algebra studied here.

\medskip

Throughout this paper we make free use of spinor conventions $\la\lambda\kappa\ra = \lambda^\al\kappa_\al=\epsilon^{\al\beta}\lambda_\beta\kappa_\al$ and similarly $[\tilde\lambda\tilde\kappa]=\tilde\lambda^\da\tilde\kappa_\da$. The twistor space $\PT$ of flat $\bbR^4$ is the total space of $\cO(1)\oplus\cO(1)\to\CP^1$. We will often use homogeneous coordinates $\lambda_\alpha$ on the $\CP^1$ base, and $\mu^\da$ on the fibres, collectively denoting these by $Z^a=(\mu^\da,\lambda_\al)$.


\section{Self-dual gravity on twistor space}
\label{sec:sdgravity-twistors}

At the classical level, self-dual Einstein gravity may be described by the twistor space action~\cite{Mason:2007ct}
\begin{equation}
\label{eq:twistor-sd-gravity-action}
S[g,h] = \int_\PT\Dif^3Z\wedge g\wedge \left(\bar\p h + \frac{1}{2}\{h,h\}\right),
\end{equation}
for fields $h\in\Omega^{0,1}(\PT,\cO(2))$ and $g\in \Omega^{0,1}(\PT,\cO(-6))$. This action is a form of holomorphic BF theory, defined with the help of the holomophic (3,0)-form
\be
    \Dif^3Z = \frac{1}{4!}\epsilon_{abcd}Z^a\dif Z^b\wedge\dif Z^c\wedge\dif Z^d = \frac{1}{8}\la\lambda\,\dif\lambda\ra\wedge[\dif\mu\wedge\dif\mu]
\ee
on $\PT$. The vertex involves a holomorphic Poisson bracket, twisted by $\cO(-2)$, that is defined by
\begin{equation}
    \{f, g\} = \epsilon^{\db\da}\cL_\da f \wedge\cL_\db g
\end{equation}
for any $(p,q)$-forms $f$, $g$ on $\PT$, where $\cL_{\dot\alpha}$ denotes the Lie derivative along $\p/\p\mu^{\dot\alpha}$. This Poisson bracket is dual to the holomorphic $(2,0)$-form
$[\dif\mu\wedge\dif\mu]/2\in\Omega^{2,0}(\PT,\cO(2))$
that plays the role of a holomorphic symplectic form on the fibres of the projection $\PT\to\CP^1$.

The action~\eqref{eq:twistor-sd-gravity-action} is invariant under Hamiltonian diffeomorphisms of twistor space, acting as
\begin{subequations}
\begin{equation}
    \label{eq:Ham-diffs-twistor}
    \delta h = \bar\p\chi + \{h,\chi\}\,,\qquad\qquad \delta g = \{g,\chi\}
\end{equation}
for a smooth function $\chi\in\Omega^0(\PT,\cO(2))$, as well as the transformations
\begin{equation}
\label{eq:g-field-trans-twistor}
\delta g = \bar\p\xi +\{h,\xi\}\,,\qquad\qquad\delta h=0\,,
\end{equation}
\end{subequations}
familiar for BF-type theories, here with $\xi\in\Omega^{0}(\PT,\cO(-6))$. In the quantum theory, twistor space diffeomorphisms are anomalous unless the theory is supersymmetrised~\cite{Mason:2007ct,Skinner:2013xp} or coupled to an axion field as in~\cite{Bittleston:2022nfr,Bittleston:2022jeq,Costello:2022wso}. In this paper, we will mostly be concerned with the classical theory.

The field equations that follow from~\eqref{eq:twistor-sd-gravity-action} are
\begin{equation}
    \bar\p h +\frac{1}{2}\{h,h\}=0\,,\qquad\qquad
    \bar\p g + \{h,g\}=0\,.
\end{equation}
The first of these asserts the integrability of the  almost complex structure $\bar\p+V$ associated to the Hamiltonian vector field $V=\{h,\ \}$. Since we deform $\PT$ using a Hamiltonian $V$, the deformed twistor space $\cPT$ will still admit a fibration $\cPT\to\CP^1$, and $[\dif\mu\wedge\dif\mu]$ survives as a holomorphic symplectic form (twisted by $\cO(2)$) on the leaves of this fibration. By Penrose's non-linear graviton construction~\cite{Penrose:1968me,Penrose:1976js}, such twistor spaces correspond to solutions of the self-dual Einstein equations. The field $g$ then represents a massless field of helicity\footnote{In our conventions, the self-dual background can be viewed as a coherent state of gravitons of helicity $+2$.} $-2$ propagating on this self-dual background.


\section{Deforming twistor space with a defect operator}
\label{sec:deforming-twistor-space}

Inspired by the twisted holography of Costello \& Gaiotto~\cite{Costello:2018zrm}, we now consider the effect of introducing a defect into this twistor space. We choose to consider a defect that couples electrically to $g$, so take the action to be
\begin{equation}
    \label{eq:twistor-defect-action}
    S[g,h] =  \int_\PT\Dif^3Z\wedge g\wedge \left(\bar\p h + \frac{1}{2}\{h,h\}\right) - \frac{\pi^2c^2}{2} \int_{\CP^1}\la\lambda\,\dif\lambda\ra\wedge(\la\al\lambda\ra\la\lambda{\beta}\ra)^2\,  g \,,
\end{equation}
where the final term describes a defect wrapping the zero section $\mu^{\dot\alpha}=0$ of $\PT\to\CP^1$.  In this term, $c^2$ is a real coupling constant measuring the strength of coupling to the defect. Since $g$ has homogeneity $-6$, the electrical coupling requires that we pick a holomorphic function of homogeneity $4$ on the $\CP^1$ defect; we chose this to be the square of,\footnote{By a theorem of Pontecorvo~\cite{Pontecorvo:1992twi} twistor spaces admitting such a holomophic function correspond to space-times with a preferred scalar-flat K{\"a}hler metric.} $\la\al\lambda\ra\la\lambda{\beta}\ra$ where $\{|\al\ra,\,|\beta\ra\}$ are an arbitrary dyad normalized so that $\la\al\beta\ra=1$.  We will use the notation
\begin{equation}
    \label{eq:p(lambda)-def}
    c(\lambda) = c\,\la\alpha\lambda\ra\la\lambda\beta\ra\,,
\end{equation} 
for later convenience.

In the presence of this defect, the equation of motion for $h$ becomes
\begin{equation}
\label{eq:deformed}
\bar\p h + \frac{1}{2}\{h,h\} = 2\pi^2 c^2(\lambda)\,\bar\delta^2(\mu)\,,
\end{equation}
where $\int \dif \mu^{\dot0}\wedge\dif\mu^{\dot1}\wedge\bar\delta^2(\mu)=1$. The equation of motion for $g$ itself is unaffected. This sourced equation is solved by
\begin{equation}
\label{eq:h-in-terms-of-phi}
 h = \frac{c^2(\lambda)}{2} \frac{[\hat\mu\,\dif\hat\mu]}{\,[\mu\,\hat\mu]^2} \,.
\end{equation}
To see this, first notice that
\begin{equation}
\label{eq:Poisson-phi}
\{h, \ \,\} = -c^2(\lambda) \,\frac{[\hat\mu\,\dif\hat\mu]}{\,[\mu\,\hat\mu]^3} \,\hat{\mu}^\da\cL_{\dot\alpha}
\end{equation}
so that $\{h,h\}=0$, both because $[\hat\mu \,\dif\hat\mu]\wedge[\hat\mu\,\dif\hat\mu]=0$ and because $\hat\mu^{\da}\cL_\da h= 0$. Now let
 \begin{equation}
\label{eq:phiEH}
\phi =  \frac{1}{4\pi^2}\frac{[\hat\mu\,\dif\hat\mu]}{\,[\mu\,\hat\mu]^2}\in\Omega^{0,1}(\PT,\cO(-2))
\end{equation}  
so $h=2\pi^2c^2(\lambda)\phi$. Then provided $[\mu\,\hat\mu]\neq0$ we have
\begin{equation}
\begin{aligned}
\bar\p\phi &= \frac{1}{4\pi^2}\left(\frac{1}{[\mu\,\hat\mu]^2}\,[\dif\hat\mu\wedge\dif\hat\mu] - 2\frac{[\mu\,\dif\hat\mu]\wedge [\hat\mu\,\dif\hat\mu]}{[\mu\,\hat\mu]^3}\right)\,\\
&= \frac{1}{4\pi^2}\left(\frac{1}{[\mu\,\hat\mu]^2}\,[\dif\hat\mu\wedge\dif\hat\mu] - \frac{[\mu\,\hat\mu]\,[\dif\hat\mu\wedge\dif\hat\mu]}{[\mu\,\hat\mu]^3}\right) =0\,.
\end{aligned}
\end{equation}
Since $[\mu\,\hat\mu]=|\mu^{\dot{0}}|^2 + |\mu^{\dot1}|^2$,
we see that $\bar\p h + \frac{1}{2}\{h,h\}$ vanishes away from the zero section $\mu^\da=0$. Furthermore, since $\phi$ is a $(0,1)$-form of homogeneity $-2$, we must have $\bar\p\phi \propto \bar\delta^2(\mu)$. We can fix the normalization by integrating $\phi$ over the $S^3$ given by $[\mu\,\hat\mu]=1$ at constant $\lambda$. On this sphere we have
\begin{equation}
\begin{aligned}
  \frac{1}{2}\int_{S^3}[\dif\mu\wedge\dif\mu]\wedge\phi 
  &= \frac{1}{8\pi^2}\int_{S^3}[\dif\mu\wedge\dif\mu]\wedge [\hat\mu\,\dif\hat\mu] 
  \\
  &= \frac{1}{8\pi^2}\int_{B}[\dif\mu\wedge\dif\mu]\wedge[\dif\hat\mu\wedge\dif\hat\mu]
  = \frac{2}{\pi^2}\,\mathrm{Vol}(B) = 1\,.
\end{aligned} 
\end{equation}
where $B$ is a unit 4-ball in the $\mathbb{C}^2$ fibre. Thus $\phi$  is correctly normalized to obey $\bar\p\phi=\bar\delta^2(\mu)$, while the corresponding $h$ obeys~\eqref{eq:deformed}.

\medskip

Let us remark that the field~\eqref{eq:phiEH} is essentially identical to the one used in section 4 of~\cite{Costello:2018zrm} in the context of twisted holography for the topological string. More precisely, the relation is
\begin{equation}
    \eta = \frac{1}{2}\phi\wedge[\dif\mu\wedge\dif\mu]\in \Omega^{2,1}_\mathrm{cl}(\PT)\,,
\end{equation}
where $\eta$ is the closed string field of the topological B model, and $\phi$ is as above. In that context, $\eta$ is related to a Beltrami differential by $\eta = V\ip \Omega$, where $\Omega$ is the holomorphic (3,0)-form on $\bbC^3$. There, the backreacted geometry sourced by the defect deforms $\bbC^3$ to the deformed conifold, viewed as $\gSL(2,\bbC)\cong\mathrm{AdS}_3\times S^3$. Similarly, the same $\phi$ plays an important role in the asymptotically flat holography of~\cite{Costello:2022jpg}, where it deforms $\PT$ to the twistor space of Burns space~\cite{Burns:1986,LeBrun:1991}, a particular scalar-flat K{\"a}hler manifold. However, in our case the relation between the Beltrami differential and the field $\phi$ is different: instead of the B-model relation $\eta = V\ip \Omega$, we have $V = \{h,\ \} = 2\pi^2c^2(\lambda)\,\{\phi,\ \}$ using the Poisson structure. In addition, here the defect acts as a source for $\bar\p h+\frac{1}{2}\{h,h\}$ rather than $\p^{-1}\eta$. These differences mean the deformed twistor space we obtain will be very different, in particular corresponding to the twistor space of a self-dual Ricci-flat manifold.

\medskip

In~\cite{Costello:2018zrm,Costello:2022jpg}, the defect was interpreted as a D1 brane of the topological B model, wrapping a holomorphic curve inside either the conifold or $\PT$. This D1 brane supports a chiral algebra that is holographically dual to the bulk theory. It would clearly be very interesting to find a string theory realisation of the bulk theory and defect of the present paper.


\subsection{The Eguchi-Hanson twistor space}
\label{sec:EH-twistor}

We'll now show that the solution~\eqref{eq:h-in-terms-of-phi}-\eqref{eq:phiEH} implies that the effect of the backreaction of the defect causes $\PT$ to be deformed to the twistor space of Eguchi-Hanson space. 

\medskip

We seek coordinates that are holomorphic with respect to the deformed Dolbeault operator 
\be \label{eq:nabla-bar}
\bar\nabla = \bar\p + \{h,\ \} = \bar\p - c^2(\lambda)\,\frac{[\hat\mu\,\dif\hat\mu]}{\,[\mu\,\hat\mu]^3}\,\hat{\mu}^{\dot\gamma} \cL_{\dot\gamma}\,.
\ee
By construction the $\lambda_\al$ are still holomorphic, so as mentioned above our curved twistor space $\mathcal{PT}$ admits a holomorphic fibration over $\CP^1$. However, the coordinates $\mu^{\da}$ are no longer holomorphic. 

Instead, consider the three functions 
\be \label{eq:newholomorphic}
X^{\da\dot{\beta}} = X^{(\da\dot\beta)} = \mu^{\da}\mu^{\dot\beta} - c^2(\lambda)\,\frac{\hat{\mu}^{\da}\hat{\mu}^{\dot\beta}}{[\mu\,\hat\mu]^2} \,,
\ee
each of homogeneity $+2$. We find
\bea
\bar\nabla X^{\da\dot\beta} &= -c^2(\lambda)\,\bar\p\left(\frac{\hat{\mu}^{\da}\hat{\mu}^{\dot\beta}}{[\mu\,\hat\mu]^2}\right) - c^2(\lambda)\,\frac{[\hat\mu\,\dif\hat\mu]}{\,[\mu\,\hat\mu]^3}\left(\mu^\da\hat\mu^{\dot\beta}+\hat{\mu}^{\da}\mu^{\dot\beta}\right) \\
&= - c^2(\lambda)\,[\mu\,\hat\mu]^3\left[ \left([\mu\,\hat\mu]\,\dif\hat{\mu}^\da -[\mu\,\dif\hat{\mu}] \,\hat{\mu}^\da +[\hat\mu\,\dif\hat\mu]\,\mu^\da\,\right)\hat{\mu}^{\dot\beta} + (\al\leftrightarrow\beta)\,\right] \\
&=0
\eea
as a consequence of the Schouten identity. Thus we can take $(X^{\da\dot\beta},\lambda_\al)$ as holomorphic coordinates on the deformed twistor space. These coordinates are not all independent, but are subject to the scaling relations $(r^2 X,r\lambda)\sim(X,\lambda)$ for $r\in\mathbb{C}^*$, as well as the constraints
\begin{equation}
\label{eq:def-twistor-X-coords}
    X^{\da\dot\beta}X_{\da\dot\beta} = -2c^2(\lambda)\,.
\end{equation}
Equivalently, setting $X^{\dot{0}\dot{0}} =X$, $X^{\dot{1}\dot{1}}=Y$ and $X^{\dot{0}\dot{1}} = X^{\dot{1}\dot{0}} = Z$, this may be written as 
\begin{equation}
\label{eq:twistor-constraint}
    XY = \left(Z-c(\lambda)\right)\left(Z+c(\lambda)\right)\,.
\end{equation}
In other words, the backreaction deforms the twistor space to the subvariety of the total space of 
\[
\cO(2)\oplus\cO(2)\oplus\cO(2)\to\CP^1
\] 
defined by these equations. This is essentially\footnote{$c(\lambda)$ vanishes when $\lambda=\al$ or $\lambda=\beta$, so these two fibres remain singular. A more precise description of the twistor space of Eguchi-Hanson space involves resolving these remaining singularities by blowing up a $\CP^1$ in each of these two fibres. See~\cite{Hitchin:1979rts} for more details.}  the twistor space of Eguchi-Hanson space~\cite{Eguchi:1978xp}. The novelty here is that we have obtained it as the backreaction of the action~\eqref{eq:twistor-sd-gravity-action} to the presence of a defect inserted in flat twistor space $\PT$.


\subsection{The space-time metric}
\label{subsec:EH-space-time}

Here we briefly review how the twistor space above corresponds to Eguchi-Hanson space-time; see \emph{e.g.}~\cite{Hitchin:1979rts,Burnett:1979st,Sparling:1981nk,Tod:1982mmp} for further details.

Recall that complexified space-time $\cM_\bbC$ arises as the space of holomorphic sections of $\cPT\to\CP^1$, \emph{i.e.}, holomorphic sections of $\cO(2)\oplus\cO(2)\oplus\cO(2)\to\CP^1$ obeying the  constraint~\eqref{eq:twistor-constraint}. For example, we may describe a section by the incidence relations
\be
\label{eq:incidence}
X^{\dot\al\dot\beta} = x^{\dot\alpha\alpha}x^{\dot\beta\beta}\left(\lambda_\alpha\lambda_\beta - \frac{4c^2\la\al\lambda\ra^2}{x^4} {\beta}_\alpha{\beta}_\beta\right)
\ee
where $x^{\da\al}$ will turn out to be Kerr-Schild coordinates on Eguchi-Hanson space\footnote{The incidence relation \eqref{eq:incidence} breaks the symmetry $\al\leftrightarrow{\beta}$. (We could of course have chosen to break it in the other direction.) 
It is possible to write incidence relations
\[
X^{\dot\al\dot\beta} = 2\im c\frac{x^{\dot\alpha\alpha}x^{\dot\beta\beta}}{x^2} \left(\al_\al\al_\beta\la\beta\lambda\ra^2 + \beta_\al\beta_\beta\la\al\lambda\ra^2\right)
\]
that obey the hypersurface constraint and preserve this symmetry, but the flat space limit is less obvious here. For this reason, we prefer to use the symmetry breaking case~\eqref{eq:incidence}. One consequence is that, even if we choose $|\beta\ra=|\hat\al\ra$ as is natural in Euclidean signature, the real Euclidean structure of the metric is not manifest in Kerr-Schild form. Instead, we can give it a natural ultrahyperbolic structure if we take the spinors $\alpha,\beta$ to each be real.}. The incidence relations~\eqref{eq:incidence} are easily verified to obey $X^{\dot\al\dot\beta}X_{\dot\al\dot\beta} = -2c^2(\lambda)$. In addition, we see that as either the defect coupling $c\to0$ or as $|x|\to\infty$ they reduce to the flat space incidence relations  $\mu^\da\mu^\db = x^{\da\al}x^{\db\beta}\lambda_\al\lambda_\beta$ for the orbifold $\bbR^4/\bbZ_2$.  It will sometimes be useful to write~\eqref{eq:incidence} as $X^{\da\db} = M^{(\da}_+(\lambda)\,M_-^{\db)}(\lambda)$ in terms of 
\be
M^{\da}_\pm(\lambda) = x^{\da\al}\lambda_\al \pm \frac{2c\la\al\lambda\ra}{x^2}\,x^{\da\al}\beta_\al\,.
\ee

The incidence relation describes a holomorphic curve $\CP^1\subset\cPT$ and the conformal structure of the corresponding space-time is given by declaring that two points $x,y\in\cM_{\bbC}$ are null separated iff the corresponding curves $\CP^1_x$ \& $\CP^1_y$ intersect. To fix the scale, we also need the $\cO(2)$-valued symplectic structure on the fibres of $\cPT\to\CP^1$. In terms of the coordinates $(X,Y,Z)$ on the fibres of $\cPT\to\CP^1$ introduced in~\eqref{eq:def-twistor-X-coords}, this is given by
\bea \label{eq:symplectic1}
\omega=\frac{1}{2\pi\im}\oint \frac{\dif X\wedge\dif Y\wedge\dif Z}{2(XY-Z^2 + c^2(\lambda))} \ =\  
\begin{cases}
 \displaystyle{\frac{\dif X\wedge\dif Z}{2X}}\qquad &\text{if $X\
\neq0$}\\[2ex]
 \displaystyle{-\frac{\dif Y\wedge \dif Z}{2Y}} \qquad &\text{if $Y\neq0$\,,}
\end{cases}
\eea
where the normalization is chosen to agree with the corresponding form $\dif\mu^{\dot0}\wedge\dif\mu^{\dot1}$ on $\PT/\bbZ_2$. More globally, this 2-form may be written  as
\be
    \label{eq:symplectic2}
    \omega = \frac{1}{8c^2(\lambda)}X^{\da\db}\,\dif X_{\dc\da}\wedge\dif X^\dc_{~\,\db}
\ee
where the $X^{\da\db}$ obey~\eqref{eq:twistor-constraint}. It is worth emphasising that this $\omega$ is still just $\dif \mu^{\dot0}\wedge \dif\mu^{\dot1}$; indeed, this follows by construction from the fact that $\cPT$ was obtained via Hamiltonian deformation. However, the $\mu^\da$ are no longer holomorphic on $\cPT$, so it is more useful to write $\omega$ in terms of the coordinates $X^{\da\db}$. The Gindikin 2-form $\Sigma(\lambda)$ is defined to be the pullback of $\omega$ to (the conormal bundle of) a $\CP^1$~\eqref{eq:incidence}.  A short calculation shows that this is
\bea
    &\Sigma(\lambda) = \frac{[M_-\dif M_-]\wedge[M_+\dif M_+]}{4c(\lambda)}\\
    &= \frac{\la\lambda\beta\ra^2}{2}[\dif u\wedge\dif u] + \la\al\lambda\ra\la\lambda\beta\ra[\dif u\wedge\dif\tilde  u] + \frac{\la\alpha\lambda\ra^2}{2}[\dif\tilde u\wedge\dif\tilde u] + \frac{c^2\la\al\lambda\ra^2}{[u\,\ut]^3}[\tilde u\,\dif u]\wedge[\tilde u\,\dif\tilde u]\,,
\eea
where we have introduced $u^{\da}=x^{\da\al}\al_\al$ and $\tilde{u}^{\da}=x^{\da\al}{\beta}_\al$. The Gindikin 2-form can be written as $\Sigma(\lambda) = e^{\da\al}\wedge e_{\da}^{\ \beta}\lambda_\al\lambda_\beta/2$, where 
\be
\label{eq:viebein-def}
    e^{\da\al} = \dif x^{\da\al} - \frac{c^2\ut^\da\al^\alpha[\ut\,\dif u]}{[u\,\ut]^3} = \bigg(\dif\ut^\da - \frac{c^2\ut^\da[\ut\,\dif u]}{[u\,\ut]^3}\bigg)\alpha^\al - \dif u^\da\beta^\alpha
\ee
defines a space-time vierbein. The corresponding metric is
\be \label{eq:EH-metric}
\dif s^2 = e^{\da\al}\odot e_{\da\al} = 2[\dif u\odot\dif\ut] + \frac{2c^2}{[u\,\ut]^3}[\ut\,\dif u]\odot[\ut\,\dif u] = \delta + \frac{16c^2}{x^6}[\tilde u\,\dif u]^{\odot 2}\,,
\ee
where $\delta= 2[\dif u\odot\dif\ut] = \dif x^{\da\al}\odot\dif x_{\da\al}$ is the flat metric. This is the Eguchi-Hanson metric, in the Kerr-Schild coordinates\footnote{Formally, we should view this as a complexified metric on the complexified space-time with complex coordinates $u^\da,\ut^\da$. The metric is real in ultrahyperbolic signature if the dyad $\{\alpha,\beta\}$ are real spinors. In this case (or in the complexified setting) we can take the limit $\alpha\to\beta$ to obtain a metric
\[ 
\dif s^2 = \delta + \frac{16c^2}{x^6}[u\,\dif u]^{\odot 2} \]
with $2^{\rm nd}$ Pleba{\'n}ski scalar $\Theta(x) = 2c^2/x^2$.} first found by Sparling \& Tod~\cite{Sparling:1981nk,Burnett:1979st,Berman:2018hwd}. (Other coordinates would be obtained from other choices of incidence relations.) It can also be written as
\be
    \dif s^2 = \delta + \tilde\p_\da \tilde\p_\db\Theta\,\dif u^\da\odot\dif u^\db
\ee
where $\tilde\p_\da = \al^{\al}\p_{\da\al}$ and the scalar
\be
\label{eq:2nd-Plebanski-scalar}
\Theta(x) = 2c^2\left(\frac{\ut^{\dot 1}}{u^{\dot 1}}\right)^2 \frac{1}{x^2}
\ee
obeys the $2^\mathrm{nd}$ Pleba{\'n}ski equation $\Delta\Theta - \frac{1}{2}\tilde\p^\da\tilde\p^\db\Theta\,\tilde\p_\da\tilde\p_\db\Theta=0$, where $\Delta = \p^{\da\al}\p_{\da\al}$ is the flat space Laplacian.

As is well known, the Eguchi-Hanson metric as written in equation \eqref{eq:EH-metric} has a singularity at the origin, which can be rectified by identifying $u^\da\sim-u^\da$. The cost of this is that the space-time is no longer globally asymptotically Euclidean, but rather only locally so. Indeed, it is the canonical example of an ALE (asymptotically locally Euclidean) spacetime. This $\bbZ_2$ quotient plays a crucial role in our narrative: the unique non-trivial Lie algebra deformation of $\ham(\bbC^2)$ is the Weyl algebra equipped with the Moyal bracket. However, its fixed point subalgebra under the $\bbZ_2$ action is the wedge subalgebra of $w_{1+\infty}$, admitting a larger family of deformations. We shall see that one of these corresponds to deforming $\bbR^4/\bbZ_2$ to Eguchi-Hanson.


\section{Celestial chiral algebras from twistor space} \label{sec:twistor-algebra}

In~\cite{Adamo:2021lrv,Costello:2022wso}, it was shown that the celestial chiral algebra of self-dual Einstein gravity is simply the loop algebra of the Poisson algebra of holomorphic functions on the fibres of twistor space over $\CP^1$. For the twistor space $\PT$ of flat space, this is simply $\cL\ham(\bbC^2)$. On the deformed twistor space above, we will see that it is instead isomorphic to a $\mu\to\infty$ scaling limit of the deformed family of $W(\mu)$ algebras~\cite{Pope:1989sr,Pope:1991ig}. This coincides with the complexification of $\mathrm{sdiff}(S^2)$, the Lie algebra of area preserving vector fields on the sphere.

As explained above, the unique member of the $W(\mu)$ family that arises as a deformation of $\ham(\bbC^2)$ is the symplecton algebra ($\mu=-3/16$). In~\cite{Bu:2022iak} this symplecton algebra was shown to arise by making $\PT$ non-commutative. The reason our deformed $\cPT$ yields a different algebra is because it is really a deformation of the orbifold $\PT/\bbZ_2$, where the non-trivial element of $\bbZ_2$ acts on the coordinates of $\PT$ by
\be
    \label{eq:Z2-action-on-twistors}
    (\mu^\da,\lambda_\al) \mapsto (-\mu^\da,+\lambda_\al)\,.
\ee
Notice that this action is independent of the overall scalings $(\mu^\da,\lambda_\al)\to r(\mu^\da,\lambda_\al)$ of the homogeneous coordinates. The source term $c^2(\lambda)\,\bar\delta^2(\mu)$, the corresponding twistor field $h = (c^2/2) [\hat\mu\,\dif\hat\mu]/[\mu\,\hat\mu]^2$ and the new holomorphic coordinates $X^{\da\db}$ are all invariant under this $\bbZ_2$ action. Indeed, when the coupling to the defect is turned off, these coordinates reduce to $X_0 = (\mu^{\dot0})^2$, $Y_0=(\mu^{\dot1})^2$ and $Z_0=\mu^{\dot0}\mu^{\dot1}$, and the relation $X_0Y_0=Z_0^2$ defines the twistor space $\PT/\bbZ_2$ corresponding to the space-time orbifold $\bbR^4/\bbZ_2$.

Only the generators of $\ham(\bbC^2)$ that are invariant under this action descend to holomorphic functions on $\PT/\bbZ_2$: these are
\bea
\label{eq:orbifold-generators}
    w[2r,2s] &= (\mu^{\dot0})^{2r}(\mu^{\dot1})^{2s} = X_0^r\,Y_0^s\\
    w[2r+1,2s+1] &= (\mu^{\dot0})^{2r+1}(\mu^{\dot1})^{2s+1} = X_0^r\,Y_0^s\, Z_0
\eea
with $r,s\in\mathbb{N}_0$. These generators transform in integer spin representations of the $\fsl_2(\bbC)$ acting on dotted spinors. Their Poisson algebra (with respect to the usual Poisson bracket $\{f,g\}=\p^\da \!f\,\p_\da g$) is exactly the wedge subalgebra of $w_{1+\infty}$, or equivalently $\ham(\bbC^2)^{\bbZ_2}$. It is thus the fact that only half the generators of the celestial chiral algebra of self-dual gravity on flat space descend to the orbifold that permits us to obtain new deformations of this algebra.


\subsection{CCA for self-dual gravity}
\label{subsec:twistor-CCA-SDG}

In this section we compute the structure constants of the Poisson algebra (\emph{i.e.} the Lie algebra of polynomial functions under the Poisson-bracket) on a fibre $\mathcal{M}_{\lambda}$ of twistor space of the Eguchi-Hanson space for a generic $\lambda \in \mathbb{CP}^1$. We will then prove that it is isomorphic to a certain $\mu \rightarrow \infty$ scaling limit of the deformed family of $W(\mu)$-algebras \cite{Pope:1989sr}. Following the arguments of~\cite{Adamo:2021lrv}, the corresponding loop algebra of it is to be identified as the celestial chiral algebra of self-dual Einstein gravity on an Eguchi-Hanson background. We will verify this by an explicit space-time calculation in section~\ref{sec:space-time-algebra}.

\medskip

The weight $-2$ Poisson bracket associated to the $\cO(2)$-valued (2,0)-form~\eqref{eq:symplectic1} is
\be\label{eq:deformed-Poisson}
\{f,g\} = 2X\left(\frac{\p f}{\p X}\frac{\p g}{\p Z} - \frac{\p f}{\p Z}\frac{\p g}{\p X}\right)
\ee
Acting on the basic coordinates, this Poisson bracket gives 
\be \label{eq:Poisson-bracket}   
 \{X,Y\} = 4Z\,,\qquad \{X,Z\} = 2X\,,\qquad \{Y,Z\} = -2Y\,,
\ee
which are just the defining relations of $\fsl_2$. Then the constraint $XY - Z^2 =- c^2(\lambda)$ is just the statement that the quadratic Casimir of this $\fsl_2$ takes the value $c^2(\lambda)$. In particular, these relations imply that the ideal 
\be
\label{eq:twistor-ideal}
\scrI = \mathrm{span}\{XY - Z^2 + c^2(\lambda)\}
\ee
is a Poisson ideal, in the sense that $\{\cO,\scrI\}\subset\scrI$.  Because of this, the Poisson bracket can straightforwardly be extended to act on the full coordinate ring
\be 
\label{eq:EHCoordinateRing}
\cO_{\cM_{\lambda}} = \bbC[X,Y,Z]\,\big/\, \mathscr{I}\,.
\ee
A natural choice of basis for this ring is
\be 
\label{eq:EH-twistor-basis}
V[2p,2q] =  X^pY^q\,,\qquad\qquad V[2p+1,2q+1] = X^pY^qZ 
\ee
with $p,q\in\mathbb{N}_0$, because any polynomial involving higher powers of $Z$ can be traded for these generators using the ideal. We emphasise the states $V[m,n]$ are defined only when $m+n\equiv0\Mod{2}$. This basis is a  deformation of the basis~\eqref{eq:orbifold-generators} to the deformed twistor space. 

 In the basis~\eqref{eq:EH-twistor-basis} we find the algebra 
\begin{subequations} \label{eqs:algebra-in-twistor-basis}
\begin{align}
\begin{split} \label{eq:alg-in-twistor-basis-1}
    &\big[V[2p,2q]\,,\,V[2r,2s]\big] = 4(ps-qr)\,V[2(p\!+\!r\!-\!1)\!+\!1,2(q\!+\!s\!-\!1)\!+\!1]\,,
\end{split} \\[2ex]
\begin{split} \label{eq:alg-in-twistor-basis-2}
&\big[V[2p,2q]\,,\,V[2r\!+\!1,2s\!+\!1]\big] = 2(p(2s+1)-q(2r+1))\,V[2(p\!+\!r),2(q\!+\!s)] \\
    &\hspace{5cm} + 4c^2(\lambda)\,(ps-qr)\,V[2(p\!+\!r\!-\!1),2(q\!+\!s\!-\!1)]\,,
\end{split} \\[2ex]
\begin{split} \label{eq:alg-in-twistor-basis-3}
 &\big[V[2p\!+\!1,2q\!+\!1]\,,\,V[2r\!+\!1,2s\!+\!1]\big]  = ((2p\!+\!1)(2s\!+\!1)-(2q\!+\!1)(2r\!+\!1))\,V[2(p\!+\!r)\!+\!1,2(q\!+\!s)\!+\!1] \\
    &\hspace{6cm} + 4c^2(\lambda)\,(ps-qr)\,V[2(p\!+\!r)\!-\!1,2(q\!+\!s)\!-\!1]\,.
\end{split}
\end{align}
\end{subequations}
The terms proportional to $c^2(\lambda)$ in~\eqref{eq:alg-in-twistor-basis-2} \&~\eqref{eq:alg-in-twistor-basis-3} are not present in $w_{1+\infty}$. These terms are easily seen to represent a non-trivial element in Lie algebra cohomology: the deformation preserves the $\fsl_2$ subalgebra, so any redefinition undoing it must act as an intertwiner. Since  $\ham(\bbC^2)^{\bbZ_2}$ decomposes into a direct sum over distinct integer spin representations, the only $\fsl_2$ equivariant redefinitions are rescalings of the generators. It's clear that such rescalings cannot trivialize the deformation \eqref{eqs:algebra-in-twistor-basis}. We identify this deformed algebra below.


\subsection{CCA in the scattering basis} \label{subsec:states-propagators}

While~\eqref{eq:EH-twistor-basis} is probably the simplest basis of $\cO_{\cM_\lambda}$, it will be helpful to consider a different basis so as to make contact with ideas of celestial holography. There, one considers the generators as conformally soft modes of the graviton~\cite{Guevara:2021abz,Strominger:2021lvk}. In flat $\bbR^4$, these are Mellin modes of momentum eigenstates, representing the external states of a scattering process.

\medskip

In our context, we should look for scattering states defined on the background deformed twistor space $\cPT$, corresponding to linearized fluctuations of the graviton around the Eguchi-Hanson background. We seek solutions that look asymptotically like plane waves. These may be represented on twistor space by 
\begin{subequations}
\label{eq:twistor-scattering-states}
\begin{equation}
\label{twistorscatteringstate}
\delta h(X,\lambda) = \int \frac{\dif t}{t^3}  \,\bar\delta^2(t|\lambda\ra - |\kappa\ra) \,\cos\left(t\sqrt{ -[\tilde\kappa|X|\tilde\kappa]}\right)
\end{equation}
for the helicity $+2$ graviton and 
\begin{equation}
\delta g(X,\lambda) = \int\frac{\dif t}{t}\, t^6 \,\bar\delta^2(t|\lambda\ra-|\kappa\ra)\,\cos\left(t\sqrt{ -[\tilde\kappa|X|\tilde\kappa]}\right)
\end{equation}
\end{subequations}
for the helicity $-2$ graviton \cite{Adamo:2022mev}, where $|\kappa\ra$ and $|\tilde\kappa]$ are fixed spinors and $[\tilde\kappa|X|\tilde\kappa] = -X^{\da\db}\tilde\kappa_{\da}\tilde\kappa_\db$. 

On space-time, the fluctuation in the $2^{\rm nd}$ Pleba{\'n}ski scalar may be obtained from $\delta h$ via the Penrose transform\footnote{Extracting the fluctuation in the Pleba{\'n}ski scalar from $\delta h\in H^{0,1}(\cPT,\cO(2))$ makes use of the twistor recursion operator for the hyperk{\"a}hler hierarchy. See~\cite{Dunajski:2000iq} for details.}
\bea
\label{eq:twistor-scatttering-state-scalar}
    \delta\Theta(x) &= \int_{\CP^1_x} \frac{\la\lambda\,\dif\lambda\ra}{\la\al\lambda\ra^4} \wedge \delta h\\
    &=\frac{1}{\la\al\kappa\ra^4}\cos\left(\sqrt{(k\cdot x)^2 - \frac{4c^2\la\al| kx|{\beta}\ra^2}{x^4}}\right)
\eea
where $k=|\kappa\ra[\tilde\kappa|$ and we have used the incidence relations~\eqref{eq:incidence}. Although it follows from the twistor construction, one can easily verify directly that this state indeed obeys the $2^{\rm nd}$ Pleba{\'n}ski equation, linearized around the Eguchi-Hanson background. Note that at asymptotically large distances  $|x|\to\infty$ in space-time, it approaches $\cos(k\cdot x)$, which is a `momentum eigenstate' on the orbifold $\bbR^4/\bbZ_2$. 

\medskip

The celestial chiral algebra is usually written in terms of the soft modes of the scattering states of the graviton. Concretely, these are the coefficients $W[p,q]$ of $(-)^{(p+q)/2}\lt_\dzero^p\lt_\done^q/p!q!$ in the expansion 
\be \label{eq:cos-Taylor-expand}
\cos\sqrt{-t[\lt|X|\lt]} = \sum_{m=0}^\infty\frac{t^{2m}[\lt|X|\lt]^m}{(2m)!}
\ee
in the twistor space scattering state. Trinomially expanding $[\lt|X|\lt]^m = (-)^m(X\lt_{\dot0}^2+Y\lt_{\dot1}^2+2Z\lt_{\dot0}\lt_{\dot1})^m$ gives
\bea
\label{eq:softbasis}
&\frac{[\lt|X|\lt]^m}{(2m)!}=  \frac{(-)^m}{(2m)!} \sum_{i+j+l=m} \binom{m}{i,j,l} X^iY^j (2Z)^l\, \Tilde{\lambda}_{\dot{0}}^{2i+l}\Tilde{\lambda}_{\dot{1}}^{2j+l}\\
&= \sum_{p+q=m}\left(\frac{(-)^{p+q}}{(2p+2q)!}\sum_{\ell=0}^{\min(p,q)}\binom{p+q}{p-\ell,q-\ell,2\ell}X^{p-\ell}Y^{q-\ell}(2Z)^{2\ell}\right)\Tilde{\lambda}_{\dot{0}}^{2p}\Tilde{\lambda}_{\dot{1}}^{2q}\\
&+ \sum_{p+q=m-1}\left(\frac{(-)^{p+q+1}}{(2p\!+\!2q\!+\!2)!}\sum_{\ell=0}^{\min(p,q)}\binom{p\!+\!q\!+\!1}{p\!-\!\ell,q\!-\!\ell,2\ell\!+\!1}X^{p-\ell}Y^{q-\ell}(2Z)^{2\ell+1}\right)\Tilde{\lambda}_{\dot{0}}^{2p+1}\Tilde{\lambda}_{\dot{1}}^{2q+1}\\
&= \sum_{p+q=m}W[2p,2q] \frac{(-)^{p+q}\Tilde{\lambda}_{\dot{0}}^{2p}\Tilde{\lambda}_{\dot{1}}^{2q}}{(2p)!(2q)!}\ \ +\sum_{p+q=m-1} W[2p+1,2q+1] \frac{(-)^{p+q+1}\Tilde{\lambda}_{\dot{0}}^{2p+1}\Tilde{\lambda}_{\dot{1}}^{2q+1}}{(2p+1)!(2q+1)!}\,,
\eea
where we have defined the generators in the scattering basis
\bea \label{eq:W-even-def}
    W[2p,2q]  &=\frac{(2p)!\,(2q)!}{(2p+2q)!}\sum_{\ell=0}^{\min(p,q)}\binom{p+q}{p-\ell,q-\ell,2\ell}X^{p-\ell}Y^{q-\ell}(2Z)^{2\ell} \\
    W[2p\!+\!1,2q\!+\!1] & = \frac{(2p+1)!\,(2q+1)!}{(2p+2q+2)!}\sum_{\ell=0}^{\min(p,q)}\binom{p\!+\!q\!+\!1}{p\!-\!\ell,q\!-\!\ell,2\ell\!+\!1}X^{p-\ell}Y^{q-\ell}(2Z)^{2\ell+1}\,.
\eea
The terms with $\ell>0$ involve higher powers of $Z$, so may be traded for powers of $X,Y$ and $c^2(\lambda)$ using~\eqref{eq:twistor-ideal}. Doing so, we find that this scattering basis is related to the basis~\eqref{eq:EH-twistor-basis} by an upper triangular transformation of the form
\begin{subequations}
\label{eq:basis-transformation}
\bea
W[2p,2q] &= \sum_{\ell=0}^{\min(p,q)}(2c(\lambda))^{2\ell} \,C_0(p,q,\ell) \,V[2p\!-\!2\ell,2q\!-\!2\ell]\,, \\
W[2p\!+\!1,2q\!+\!1] &= \sum_{\ell=0}^{\min(p,q)}(2c(\lambda))^{2\ell}\,C_1(p,q,\ell)\, V[2p\!-\!2\ell\!+\!1,2q\!-\!2\ell\!+\!1]\,,
\eea
with coefficients
\be
\label{eq:c0c1}
C_0(p,q,\ell)=\frac{[p]_\ell\,[q]_\ell\,[p+q]_\ell}{\ell!\,[2(p+q)]_{2\ell}}\,,\qquad\qquad
C_1(p,q,\ell)= \frac{[p]_{\ell}
\,[q]_{\ell}\,[p+q+1]_{\ell}}{\ell!\,[2(p+q+1)]_{2\ell}}\,,
\ee
\end{subequations}
where $[p]_\ell = p!/(p-\ell)!$ is the descending Pochhammer symbol. Note that the two bases coincide when the coupling $c$ to the defect is sent to zero.   

In terms of the scattering basis, the algebra takes the form
\bea
\label{eq:sl-infinity} 
&\big[W[p,q],W[r,s]\big] \\
&= \sum_{\ell\geq0} (2c(\lambda))^{2\ell}\,R_{2\ell+1}(p,q,r,s)\,\psi_{2\ell+1}\bigg(\frac{p\!+\!q}{2},\frac{r\!+\!s}{2}\bigg)\,W[p\!+\!r\!-\!2\ell\!-\!1,q\!+\!s\!-\!2\ell\!-\!1]\,, 
\eea
where 
\be
\psi_{2\ell+1}(m,n) = (-)^\ell\frac{[\ell+1/2]_\ell}{4^{2\ell}\,[m-1/2]_\ell\,[n-1/2]_\ell\,[m+n-1/2-\ell]_\ell}\,.
\ee
Comparing this to the commutation relations~\eqref{eq:W(mu)-commutator} of the $W(\mu)$-algebras, we see that the function $\Psi_{2\ell+1}(m,n;\sigma)$ in~\eqref{eq:Psi-ell} has been replaced by $\psi_{2\ell+1}(m,n)$. It's instructive to compare these functions in the case $\ell=1$. We have
\be \label{eq:Psi-3} 
\Psi_3(m,n;\sigma) = 1 - \frac{3(4\sigma+1)(4\sigma+3)}{(2m-1)(2n-1)(2(m+n)-3)}\,. \ee
It's then clear that in the scaling limit $\sigma\to\infty$, $\fq\to0$ with $4\fq\sigma = c(\lambda)$ held fixed
\be 
\lim_{{\sigma\to\infty}\atop{\fq\to0}}\fq^2\Psi_3(m,n;\sigma) = - \frac{3c^2(\lambda)}{(2m-1)(2n-1)(2(m+n)-3)} = 4c^2(\lambda)\,\psi_3(m,n)\,. 
\ee
Similarly, 
\be
\lim_{{\sigma\to\infty}\atop{\fq\to0}}\fq^{2\ell}\Psi_{2\ell+1}(m,n;\sigma) =  (2c(\lambda))^{2\ell}\,\psi_{2\ell+1}(m,n)\,.
\ee
It follows from the fact that the scattering and twistor bases are related by the (invertible) upper triangular transformation~\eqref{eq:basis-transformation} that the algebras~\eqref{eqs:algebra-in-twistor-basis} \&~\eqref{eq:sl-infinity} are isomorphic. (In appendix \ref{app:isomorphism} we check explicitly that this transformation respects the Lie brackets, see also~\cite{Pope:1989sr,Bergshoeff:1989ns}). The algebras on different generic,\footnote{In the twistor space defined by~\eqref{eq:twistor-constraint} the fibres $\lambda=\al$ and $\lambda=\beta$ remain singular. In the true twistor space of Eguchi-Hanson space-time these singularities are resolved by blowing up $T^*\CP^1\to\bbC^2/\bbZ_2$ \cite{Hitchin:1979rts}. It's nonetheless easy to verify that the Poisson algebra of global regular functions on $T^*\CP^1$ (in its standard complex structure) remains isomorphic to $w_\wedge$.} fibres $\cM_\lambda$ are all isomorphic, as may be seen by rescaling the generators by an appropriate power of $c(\lambda)$. We call this algebra
\be
    W(\infty) = \lim_{{\fq\to0}\atop{\mu\to\infty}} W(\mu)\,,\qquad\fq\sqrt{\mu} \quad\text{fixed}\,.
\ee

\medskip

The CCA of self-dual gravity on the Eguchi-Hanson background therefore has OPEs
\bea
&W[p,q](\lambda_1)\,W[r,s](\lambda_2) \\
&\sim - \frac{\tau^{p+q-3}}{2\la12\ra}\sum_{\ell\geq0} (2c(\lambda_2))^{2\ell}\,R_{2\ell+1}(p,q,r,s)\,\psi_{2\ell+1}\bigg(\frac{p\!+\!q}{2},\frac{r\!+\!s}{2}\bigg)\,W[p\!+\!r\!-\!2\ell\!-\!1,q\!+\!s\!-\!2\ell\!-\!1](\lambda_2)\,,
\eea
where $\tau = \la\al1\ra/\la\al2\ra$ has been introduced to give both sides the appropriate weight. (Note that $\tau$ is independent of $\alpha$ working modulo $\la12\ra$.) In the inhomogeneous coordinates $z_i = \la\al i\ra/\la i\beta\ra$ this reads\footnote{The soft modes $W[p,q](\lambda)$ should be viewed as sections of $\cO(p+q-4)$ over the celestial sphere. Working inhomogeneously in terms of $z = \la\alpha\kappa\ra/\la\kappa\beta\ra$ these coincide with the modes obtained via Mellin transform.}
\bea
&W[p,q](z_1)\,W[r,s](z_2) \\
&\sim - \frac{1}{2z_{12}}\sum_{\ell\geq0} (2cz_2)^{2\ell}\,R_{2\ell+1}(p,q,r,s)\,\psi_{2\ell+1}\bigg(\frac{p\!+\!q}{2},\frac{r\!+\!s}{2}\bigg)\,W[p\!+\!r\!-\!2\ell\!-\!1,q\!+\!s\!-\!2\ell\!-\!1](z_2)\,. \eea
Decomposing $W[p,q](z)$ into Laurent modes in $z$ we can see the CCA is isomorphic to $\cL W(\infty)$\footnote{It's intriguing that this isomorphism requires rescaling generators in a $z$ dependent way. We expect this modifies the modules at $z=0,\infty$ determining the vertex algebra vacua.}.


\subsection{CCA for self-dual Yang-Mills}
\label{subsec:twistor-CCA-SDYM}

It is straightforward to extend these considerations to the celestial chiral algebra of self-dual Yang-Mills (for the semisimple gauge algebra $\fg$ with invariant bilinear form $\tr$), sometimes called the $S$-algebra. Again, this algebra is deformed when describing self-dual Yang-Mills on an Eguchi-Hanson background. Classically, this may described by the twistor space action
\be \label{eq:sdYM-twistor-action}
S[b,a]= \int_\cPT\Omega\wedge\tr(b\wedge f)\,
\ee
for the fields $b\in\Omega^{0,1}(\cPT,\cO(-4)\otimes\fg)$  and $a\in\Omega^{0,1}(\cPT,\fg)$, where $f =\bar\nabla a+\frac{1}{2}[a,a]$. The main difference compared to self-dual gravity is that the vertex now involves the Lie bracket on $\fg$ rather than the Poisson bracket.\footnote{The Dolbeault operator $\bar\nabla$ appearing in the curvature $(0,2)$-form is just the usual $\bar\p$ operator when written in terms of holomorphic coordinates $(X,Y,Z,\lambda)$ on $\cPT$.}  The deformed $S$-algebra is thus simply the loop algebra of  $\fg\otimes\cO_{\cM_\lambda}$.

\medskip

As before, the most natural choice of basis for $\fg\otimes\cO_{\cM_\lambda}$ is 
\be 
\label{eq:GaugePolynomials} I_\sfa[2p,2q] = t_\sfa X^pY^q\,,\qquad\qquad I_\sfa[2p+1,2q+1] = t_\sfa X^pY^qZ\,,
\ee
where the $t_\sfa$ form a basis of $\fg$. Note that again $I_\sfa[m,n]$ is only defined for $m+n\equiv0\Mod{2}$. The structure constants follow immediately from the coordinate ring $\cO_{\cM_\lambda}$ and are given by 
\bea \label{eq:twistor-S-structure-constants}
\big[I_\sfa[2p,2q],I_\sfb[2r,2s]\big] &= f_{\sfa\sfb}^{~~\sfc}\,I_\sfc[2p\!+\!2r,2q\!+\!2s]\,, \\
\big[I_\sfa[2p,2q],I_\sfb[2r\!+\!1,2s\!+\!1]\big] &= f_{\sfa\sfb}^{~~\sfc}\,I_\sfc[2p\!+\!2r\!+\!1,2q\!+\!2s\!+\!1]\,,\\
\big[I_\sfa[2p\!+\!1,2q\!+\!1],I_\sfb[2r\!+\!1,2s\!+\!1]\big] 
&= f_{\sfa\sfb}^{~~\sfc}\,\big(I_\sfc[2p\!+\!2r\!+\!1,2q\!+\!2s\!+\!1] + c^2(\lambda)\, I_\sfc[2p\!+\!2r,2q\!+\!2s]\big)\,.
\eea
Once again, we can change basis to the soft modes appearing in the expansion of the scattering states, which now include a color factor. Defining generators $J_\sfa[r,s]$ in the scattering basis via
\bea
\label{eq:twistor-wavefunction} 
    &t_\sfa\cos\sqrt{-[\lt|X|\lt]} = t_\sfa\sum_{m=0}^\infty\frac{[\lt|X|\lt]^m}{(2m)!}\\
    &=\sum_{p+q=m}J_\sfa[2p,2q] \frac{(-)^{p+q}\Tilde{\lambda}_{\dot{0}}^{2p}\Tilde{\lambda}_{\dot{1}}^{2q}}{(2p)!(2q)!}\ \ +\sum_{p+q=m-1} \!\!J_\sfa[2p\!+\!1,2q\!+\!1] \frac{(-)^{p+q}\Tilde{\lambda}_{\dot{0}}^{2p\!+\!1}\Tilde{\lambda}_{\dot{1}}^{2q+1}}{(2p\!+\!1)!\,(2q\!+\!1)!}\,,
\eea
the $I$ and $J$ bases are related by the same upper triangular transformation that we met in self-dual gravity:
\bea
J_\sfa[2p,2q]&=\sum_{\ell=0}^{\min(p,q)}(2c(\lambda))^{2\ell}\, C_0(p,q,\ell)\, I_\sfa[2(p-\ell),2(q-\ell)]\,,\\
J_\sfa[2p\!+\!1,2q\!+\!1]&= \sum_{\ell=0}^{\min(p,q)}(2c(\lambda))^{2\ell}\,C_1(p,q,\ell)\,I_\sfa[2(p-\ell)\!+\!1,2(q-\ell)\!+\!1]\,,
\eea
where $C_0$, $C_1$ were given in~\eqref{eq:c0c1}. In the scattering basis, the deformed $S$-algebra becomes
\bea
\label{eq:SDYM-CCA-twistor}
&[J_\sfa[p,q],J_\sfb[r,s]] = \\ &f_{\sfa\sfb}^{~~\sfc}\sum_{\ell=0}^{\infty}(2c(\lambda))^{2\ell}\, R_{2\ell}(p,q,r,s)\,\psi_{2\ell}\bigg(\frac{p+q}{2},\frac{r+s}{2}\bigg) 
J_\sfc[p+r-2\ell,q+s-2\ell]\,.
\eea
with 
\be
\psi_{2\ell}(m,n) = (-)^\ell\frac{[\ell-1/2]_\ell}{4^{2\ell}[m-1/2]_\ell[n-1/2]_\ell[m+n-1/2-\ell]_\ell}\,.
\ee
These structure constants arise as a scaling limit of a family $S_\wedge(\mu;\fq)$ of deformed $S$-algebras (in the case $\fg=\fgl(N)$). More precisely, in analogy to $W(\mu;\fq)$, we define $S_\wedge(\mu;\fq)$ by the relations
\bea \label{eq:S-sigma-algebra}
&\left[\widetilde J_\sfa[p,q],\widetilde J_\sfb[r,s]\right] = f_{\sfa\sfb}^{~~\sfc}\sum_{\ell=0}^\infty\fq^{2\ell}R_{2\ell}(p,q,r,s)\Psi_{2\ell}\bigg(\frac{p+q}{2},\frac{r+s}{2};\sigma\bigg)\widetilde J_\sfc[p+r-2\ell,q+s-2\ell] \\
&+ d_{\sfa\sfb}^{~~\sfc}\sum_{\ell=0}^\infty\fq^{2\ell+1}R_{2\ell+1}(p,q,r,s)\Psi_{2\ell+1}\bigg(\frac{p+q}{2},\frac{r+s}{2};\sigma\bigg)\widetilde J_\sfc[p+r-2\ell,q+s-2\ell]
\eea
where $\Psi_{\ell}(m,n;\sigma)$ is given in \eqref{eq:Psi-ell}. Here $d_{\sfa\sfb\sfc} = \tr(t_\sfa\{t_\sfb,t_\sfc\})$, which arises in the non-commutative setting but drops out in the scaling limit. See \cite{Monteiro:2022xwq} for further details. Sending $\sigma\to\infty$, $\fq\to0$ with $4\sigma\fq=c(\lambda)$ fixed gives
\bea
&\lim_{\sigma\to\infty}\fq^{2\ell}\Psi_{2\ell}(m,n;\sigma) = (2c(\lambda))^{2\ell}\psi_{2\ell}(m,n) \\
&= (2c(\lambda))^{2\ell}(-)^\ell\frac{[\ell-1/2]_\ell}{4^{2\ell}[m-1/2]_\ell[n-1/2]_\ell[m+n+1/2-\ell]_\ell}\,,
\eea
agreeing with \eqref{eq:SDYM-CCA-twistor} and defining a family of Lie algebras $S_\wedge(\infty;c)$. For $\lambda\neq\alpha,\beta$ and $c\neq0$ these are all isomorphic to $S_\wedge(\infty)\cong S_\wedge(\infty;1)$, so we obtain the same algebra on generic twistor fibres.

Therefore, the defining OPEs of the self-dual Yang-Mills CCA on Eguchi-Hanson are
\bea
&J_\sfa[p,q](\lambda_1)J_\sfb[r,s](\lambda_2) \\
&\sim - \frac{\tau^{p+q-1}f_{\sfa\sfb}^{~~\sfc}}{2\la12\ra}\sum_{\ell=0}^{\infty}(2c(\lambda))^{2\ell}\, R_{2\ell}(p,q,r,s)\,\psi_{2\ell}\bigg(\frac{p+q}{2},\frac{r+s}{2}\bigg) 
J_\sfc[p+r-2\ell,q+s-2\ell](\lambda_2)\,.
\eea
Or, in inhomogeneous coordinates
\bea
&J_\sfa[p,q](z_1)J_\sfb[r,s](z_2) \\
&\sim - \frac{f_{\sfa\sfb}^{~~\sfc}}{2z_{12}}\sum_{\ell=0}^{\infty}(2cz_2)^{2\ell}\, R_{2\ell}(p,q,r,s)\,\psi_{2\ell}\bigg(\frac{p+q}{2},\frac{r+s}{2}\bigg) 
J_\sfc[p+r-2\ell,q+s-2\ell](z_2)\,.
\eea


\section{Splitting functions on the Eguchi-Hanson background} \label{sec:space-time-algebra}

On space-time, self-dual gravity may be described perturbatively around a self-dual background by the Chalmers-Siegel action~\cite{Siegel:1992wd,Chalmers:1996rq} 
\begin{equation}
\label{eq:Chalmers-Siegel-SDGR-action}
    S[\Tilde{\Theta},\Theta] = \int \dif^4x\,\bigg(\p^{\da\al}\Tilde{\Theta}\,\p_{\da\al}\Theta + \frac{1}{2}\Tilde{\Theta}\, \tilde\p^\da\tilde\p^\db\Theta\,\tilde\p_\da\tilde\p_\db\Theta\bigg)\,, 
\end{equation}
where, as before, $\tilde\p_{\da}=\al^\al\p_{\da\al}$. This action is equivalent to the twistor action~\eqref{eq:twistor-sd-gravity-action} at the classical level. Varying $\Tilde{\Theta}$ leads to Pleba{\'n}ski's second heavenly equation
\begin{equation}
    \label{eq:2nd-Plebanski}
    \Delta\Theta - \frac{1}{2}\tilde\p^\da\tilde\p^\db\Theta\,\tilde\p_\da\tilde\p_\db\Theta=0\,,
\end{equation}
with $\Delta=\p^{\da\al}\p_{\da\al}$. The Laplacian $\Delta_g$ on the self-dual background defined by a solution of~\eqref{eq:2nd-Plebanski} can be written as
\be
\Delta_g = \Delta - (\tilde\p^\da\tilde\p^\db\Theta)\tilde\p_\da\tilde\p_\db\,,
\ee
so that the remaining field equation
\be
0=\Delta\Tilde{\Theta} - \{\tilde\p^\da\Theta,\tilde\p_\da\Tilde\Theta\} = \Delta_g\Tilde\Theta
\ee
reveals $\Tilde\Theta$ as representing a linearized negative helicity graviton propagating on the self-dual background. We will often write $\tilde\p^\da\tilde\p^\db\Theta\,\tilde\p_\da\tilde\p_\db\Theta= \{\tilde\p^\db\Theta,\tilde\p_\db\Theta\}$, where $\{\ , \ \}$ is a Poisson bracket on space-time which coincides with the twistor bracket on the fibre over $\lambda=\beta$.

\begin{figure}[t!]
    \centering
    \includegraphics[scale=0.33]{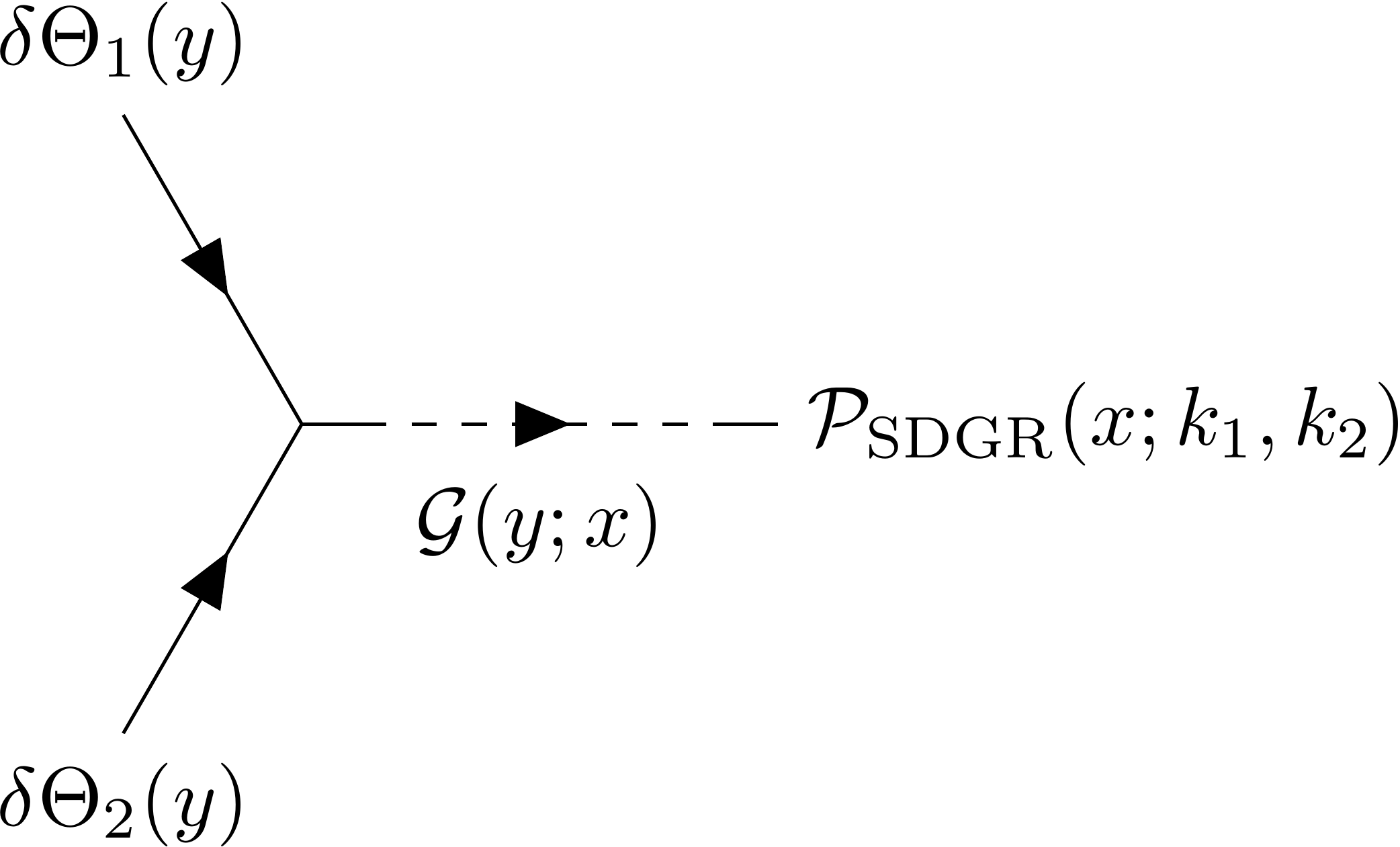}
    \caption{\emph{Tree contribution to the perturbiner in self-dual gravity. This is the position space counterpart of figure \ref{fig:splitting}, which determines the tree splitting function.}} \label{fig:perturbiner}
\end{figure}

In this section, we will recover the celestial chiral algebras obtained in section~\ref{sec:twistor-algebra}  by considering the splitting functions for positive helicity fluctuations of self-dual gravity and self-dual Yang-Mills around the Eguchi-Hanson background $\Theta=2c^2(\tilde u^{\dot1}/u^{\dot1})^2(1/x^2)$. These calculations provide an independent check of the results obtained by twistor theory above. In particular, in the gravitational case we  compute the residue of the holomorphic collinear singularity in 
\be \label{eq:SDGR-perturbiner-int} 
\cP_\mathrm{SDGR}(x;k_1,k_2) = \int_{\bbR^4/\bbZ_2}\dif^4y\,\cG(x;y)\,\{\tilde\p^\da\delta\Theta_1(y),\tilde\p_\da\delta\Theta_2(y)\}\,,
\ee
where $\delta\Theta_i(y)$ are fluctuations in the Pleba{\'n}ski scalar and $\cG(x,y)$ is the Green's function for the background Laplacian. 
$\cP_\mathrm{SDGR}$ is known as the \emph{perturbiner} of self-dual gravity~\cite{Berends:1987me,Rosly:1996vr,Rosly:1997ap}. The integral \eqref{eq:SDGR-perturbiner-int} can be obtained from the partially off-shell Feynman diagram illustrated in figure \ref{fig:perturbiner}. This is simply the position space analogue of the diagram leading to the gravitational splitting function, as depicted in figure \ref{fig:splitting}. It encodes the same information, and can be used to extract the celestial OPE.

Differentiating under the integral shows that the perturbiner obeys
\be \label{eq:perturbiner-def}
    \Delta_g \cP_\mathrm{SDGR} = \{\tilde\p^\da\delta\Theta_1,\tilde\p_\da\delta\Theta_2\}\,,
\ee
where $\Delta_g$ is the Eguchi-Hanson Laplacian. As above, we consider the fluctuations 
\be 
\label{eq:Plebanski-fluctuations}
    \delta\Theta_i(x) = \frac{1}{\la\al i\ra^4} \cos\sqrt{(k_i\cdot x)^2 - \frac{4c^2\la\alpha|k_i x|\beta\ra^2}{x^4}}
\ee
around Eguchi-Hanson space, while the scalar Green's function is~\cite{Page:1979ga,Atiyah:1981ey}
\be
\cG(x;y) = -\frac{x^2 + y^2}{2\pi^2}\Bigg((x^2+y^2)^2 - 4[u\,\vt]^2w(y) - 4[v\,\ut]^2w(x) - 8[u\,\vt][v\,\ut]\bigg(1+\frac{4c^2}{x^2y^2}\bigg)\Bigg)^{-1}\,, \ee
in the Kerr-Schild coordinates, where $u = x|\al\ra$, $\ut=x|\beta\ra$, $v=y|\al\ra$ and $\vt=y|\beta\ra$, and where we have set  $w(x) = 1 - 4c^2/x^4$.

The space-time calculation turns out to be considerably more involved than the twistor space arguments. We will content ourselves with expanding both sides of equation \eqref{eq:SDGR-perturbiner-int} in powers of $c^2$ as
\bea
&\cP_\mathrm{SDGR}(x;k_1,k_2) = \sum_{n=0}^\infty c^{2n}\,\cP_\mathrm{SDGR}^{(n)}(x;k_1,k_2)\,, \\
&\cG(x;y) = \sum_{n=0}^\infty c^{2n}\,\cG^{(n)}(x;y)\,,\qquad \delta\Theta_i(x) = \sum_{n=0}^\infty c^{2n}\,\delta\Theta_i^{(n)}(x)\,,
\eea
and working just to first non-trivial order in $c^2$.


\subsection{CCA for self-dual gravity on the orbifold} 
\label{subsec:CCA-orbifold}

At zeroth order in $c^2$ we expect to recover the fixed point subalgebra of $\cL\ham(\bbC^2)$ under $\bbZ_2$, \emph{i.e.}, the loop algebra of the wedge subalgebra of $w_{1+\infty}$. To check this, we use the zeroth order states $\delta\Theta_i^{(0)}(x) = \cos(k_i\cdot x)/\la\al i\ra^4$ and
propagator
\be
\cG^{(0)}(x;y) = - \frac{x^2+y^2}{2\pi^2(x^2-y^2)^2} = - \bigg(\frac{1}{4\pi^2(x-y)^2} + \frac{1}{4\pi^2(x+y)^2}\bigg)\,.
\ee
Strictly, in this section we're working in ultrahyperbolic signature and our propagator should be defined using an $\im\eps$ prescription. To evaluate our integrals, we Wick rotate to Euclidean signature and complexify the momenta. We leave these steps implicit for the remainder of the section. Plugging the above propagator and states into the zeroth order part of equation~\eqref{eq:SDGR-perturbiner-int} gives
\bea \label{eq:SDGR-orbifold-perturbiner}
\cP_\mathrm{SDGR}^{(0)}(x;k_1,k_2) &= - \frac{[12]^2}{\la\alpha1\ra^2\la\alpha2\ra^2}\int_{\bbR^4}\dif^4y\,\frac{\cos(y\cdot k_1)\cos(y\cdot k_2)}{4\pi^2(x-y)^2}\\
&=-\frac{[12]}{\la12\ra} \frac{\cos(x\cdot k_-)-\cos(x\cdot k_+)}{4\la\al1\ra^2\la\al2\ra^2}\,,
\eea
where $k_\pm = k_1\pm k_2$.  We see that $\cP^{(0)}_{\rm SDGR}$ is the usual gravitational splitting function $[12]/\la12\ra$ times a wavefunction on the orbifold that combines the two momentum of the original states.

In the holomorphic collinear limit, modulo non-singular terms, we can take $k_\pm = (\tau|1]\pm|2])\la2|$ and set $\tau = \la\al1\ra/\la\al2\ra$. The familiar flat space correspondence between null momentum eigenstates and hard graviton generating functions $\delta\Theta_k(x)\leftrightarrow w(\kt,\kappa)$ goes through largely unchanged, although we now require that $w(\kt,\kappa) = w(-\kt,\kappa)$. Making this identification in equation \eqref{eq:SDGR-orbifold-perturbiner} we recover the celestial OPE
\be w(\kt_1,\kappa_1)\,w(\kt_2,\kappa_2)\sim\frac{[12]}{4\la12\ra}\tau^{-2}\big(w(\tau\kt_1-\kt_2,\kappa_2) - w(\tau\kt_1+\kt_2,\kappa_2)\big)\,.
\ee
Note that the difference on the right hand side simply projects onto the $\bbZ_2$-invariant terms in both $\kt_1$ and $\kt_2$. Extracting the soft modes via Mellin transforms
\be \mathrm{Res}_{\Delta=-2m}\int_0^\infty\dif\omega\,\omega^{\Delta-1}\delta\Theta_k(x) = (-)^m\sum_{p+q=2m}\frac{\tilde z^q}{p!q!}w[p,q](z) \ee
we find that
\be \label{eq:CCA-orbifold} w[p,q](z_1)w[r,s](z_2)\sim - \frac{ps-qr}{2z_{12}}w[p\!+\!r\!-\!1,q\!+\!s\!-\!1](z_2)\,. \ee
Here we've expressed the OPE in terms of the inhomogeneous coordinates $z_i = \la\alpha\kappa_i\ra/\la\kappa_i\beta\ra$ chosen so that $\kappa_i = \beta$ lies at $z_i=\infty$. As expected, these are the defining relations of $\cL\ham(\bbC^2)^{\bbZ_2}$.


\subsection{The correction at order $c^2$} \label{subsec:simplify-first-order}

Now let's consider the first order correction to the perturbiner in $c^2$. The Laplacian on Eguchi-Hanson space is 
\be 
\Delta_g = \Delta^{(0)} + c^2\Delta^{(1)} = \Delta_\delta - \frac{16c^2\ut^\da\ut^\db}{x^6}\dt_\da\dt_\db\,, \ee
so the first order part of the perturbiner obeys 
\be \label{eq:SDGR-perturbiner-dif-1} \Delta^{(0)}\cP_\mathrm{SDGR}^{(1)}(x;k_1,k_2) = \{\tilde\p^\da\delta\Theta_1^{(0)}(x),\tilde\p_\da\delta\Theta_2^{(1)}(x)\} + (1\leftrightarrow2) - \Delta^{(1)}\cP^{(0)}_\mathrm{SDGR}(x;k_1,k_2)\,. 
\ee
We've already seen that in the holomorphic collinear singularity in $\cP^{(0)}_\mathrm{SDGR}(x;k_1,k_2)$ is a linear combination of null momentum eigenstates on flat space. The effect of the third term on the right hand side is simply to shift these null momentum eigenstates to their curved space counterparts at first order in $c^2$. Hard graviton generating functions in the CCA are identified with null momentum eigenstates on the Eguchi-Hanson background; therefore, this term does not modify the singular part of the celestial OPE.\footnote{This argument is a little too slick. It could be that the non-singular part of the flat space perturbiner $\cP^{(0)}_\mathrm{SDGR}(x;k_1,k_2)$ involves terms of the form $\la12\ra\log\la12\ra$. These can generate holomorphic collinear singularities when we differentiate under the integral sign to perform the Fourier transform. Terms of this type are in fact present, but it's not hard to show that they don't contribute to the celestial OPE.}

It's therefore sufficient to compute the holomorphic collinear singularity in
\be \label{eq:SDGR-singularity-integral} \int_{\bbR^4/\bbZ_2}\dif^4y\,\cG^{(0)}(x;y)\{\tilde\p^\da\delta\Theta_1^{(0)},\tilde\p_\da\delta\Theta_2^{(1)}\}
= - \frac{1}{4\pi^2}\int_{\bbR^4}\frac{\dif^4y}{(x-y)^2}\,\{\tilde\p^\da\delta\Theta_1^{(0)},\tilde\p_\da\delta\Theta_2^{(1)}\}\,. 
\ee
The order $c^2$ piece of the null momentum eigenstate is
\be \label{eq:first-order-eigenstate} 
\delta\Theta_i^{(1)}(y) = \frac{2}{\la\alpha i\ra^4}\bigg(\frac{\la\alpha|k_i y|\beta\ra}{y^2}\bigg)^2\frac{\sin(k_i\cdot y)}{k_i\cdot y} = \frac{2}{\la\alpha i\ra^4}\bigg(\frac{\la\al|k_i y|\beta\ra}{y^2}\bigg)^2\int_0^1\dif s\,\cos(s\,y\cdot k_i)\,, 
\ee
and so
\bea \label{eq:SDGR-vertex}
&\{\tilde\p^\da\delta\Theta_1^{(0)}(y),\tilde\p_\da\delta\Theta_2^{(1)}(y)\} \\
&= \frac{4\cos(y\cdot k_1)}{\la\alpha 1\ra^2\la\alpha 2\ra^2}\Bigg(\frac{2[\vt 1][v 2]}{y^6}\bigg([12] - \frac{6[v1][\vt 2]}{y^2}\bigg)\int_0^1\dif s\,\cos(s\,y\cdot k_2) \\
&+ \frac{[\vt 2]\la\alpha 2\ra[12]}{y^4}\bigg([12] - \frac{2([\vt 1][v2]+[v1][\vt 2])}{y^2}\bigg)\int_0^1\dif s\,s\sin(s\,y\cdot k_2) \\
&+ \frac{[\vt 2]^2\la\alpha2\ra^2[12]^2}{2y^4}\int_0^1\dif s\,s^2\cos(s\,y\cdot k_2)\Bigg)\,,
\eea
where again $|v]=y|\alpha\ra$ and $|\vt] = y|\beta\ra$.

\medskip

Given the number of terms present, evaluating the collinear singularity in equation \eqref{eq:SDGR-singularity-integral} is somewhat tedious. As such, we relegate the detailed computation to appendix \ref{app:SDGR-calcs}, and simply sketch the calculation for the final term; that is, we wish to compute the holomorphic collinear singularity in
\be -\frac{[12]^2}{2\pi^2\la\alpha1\ra^2}\int_0^1\dif s\,s^2\int_{\bbR^4}\frac{\dif^4y}{(x-y)^2y^4}\,[\vt2]^2\cos(y\cdot k_1)\cos(s\,y\cdot k_2)\,. 
\ee
The coefficient of the integrals depending only on spinor-helicity variables can be ignored for the moment. The first step is to combine the cosines using the double angle formula, and replace the $[\vt2]^2$ factor in the integrand with derivatives to get
\be \label{eq:SDGR-simplified-integral} \frac{[12]^2}{4\pi^2\la\alpha1\ra^2}\la\beta\p_{\lambda_2}\ra^2\int_0^1\dif s\,\int_{\bbR^4}\frac{\dif^4y}{(x-y)^2y^4}\,\big(\cos(y\cdot k_-(s)) + \cos(y\cdot k_+(s))\big)\,, \ee
where $k_\pm(s) = k_1\pm sk_2$. Let's consider the integral 
\be \label{eq:Im} \cI_m(x;k) = \int_{\bbR^4}\frac{\dif^4y}{(x-y)^2y^{2(m+1)}}\,\cos(y\cdot k) \ee
for $m\in\bbZ$, which when $m=1$ appears twice in the inner integral of \eqref{eq:SDGR-simplified-integral}. It suffers from a divergence of order $2(m-1)$ as $y\to0$ for $m\in\bbZ_{\geq1}$. However, after taking the derivatives with respect to $\lambda_2$ in equation \eqref{eq:SDGR-simplified-integral} will obtain a finite answer.\footnote{This is step is not strictly necessary: we can retain factors of $[vi],[\vt i]$ in the integrand, which are ultimately integrated against a Gaussian in a straightforward way. This would keep our integrals finite throughout the calculation. However, it's more computationally convenient to absorb these factors into derivatives with respect to spinor helicity variables.} We can rewrite \eqref{eq:Im} using standard tricks. First, Feynman parametrisation gives
\be \label{eq:Im-simplified} \cI_m(x;k) = (m+1)\int_0^1\dif t\,(1-t)^m\int_{\bbR^4}\frac{\dif^4y}{((y-tx)^2 + t(1-t)x^2)^{m+2}}\,\cos(y\cdot k)\,. \ee
Shifting $y\mapsto \tilde y = y+tx$
\bea &\cI_m(x;k) = (m+1)\int_0^1\dif t\,(1-t)^m\cos(t\,x\cdot k)\int_{\bbR^4}\frac{\dif^4\tilde y}{(\tilde y^2 + t(1-t)x^2)^{m+2}}\,\cos(\tilde y\cdot k) \\
&= \frac{1}{m!}\int_0^1\dif t\,(1-t)^m\cos(t\,x\cdot k)\int_0^\infty\dif r\,r^{m+1}e^{-rt(1-t)x^2}\int_{\bbR^4}\dif^4\tilde y\,\cos(\tilde y\cdot k)e^{-r\tilde y^2}\,.
\eea
The space-time integral is now a straightforward Fourier transform of a Gaussian, giving
\bea \label{eq:Im-fully-simplified}
&\cI_m(x;k) = \frac{\pi^2}{m!}\int_0^1\dif t\,(1-t)^m\cos(t\,x\cdot k)\int_0^\infty\dif r\,r^{m-1}e^{-rt(1-t)x^2 - k^2/4r} \\
&= \frac{\pi^2}{m!}\int_0^1\dif t\,(1-t)^m\cos(t\,x\cdot k)\int_0^\infty\dif r\,r^{m-1}e^{-rt(1-t)x^2-k^2/4r}\,. \eea
The $r$ integral can be performed directly by making the substitution $\nu = -r t(1-t)x^2 - k^2/4r$. We have
\be \label{eq:BesselK} \int_0^\infty\dif r\,r^{m-1}e^{-rt(1-t)x^2 - k^2/4r} = 2\bigg(\frac{k^2}{4t(1-t)x^2}\bigg)^{m/2}K_m(\sqrt{t(1-t)x^2k^2})\,, \ee
where $K_m(z)$ denotes a modified Bessel function of the second kind. For $m\in\bbZ_{\geq1}$ this has a pole of order $m$ in $t(1-t)x^2$, leading to a divergence in the outer integral of \eqref{eq:Im-fully-simplified} as $t\to0$. This reflects the divergence in the original expression \eqref{eq:Im} as $y\to0$. For $m\in\bbZ_{\leq-1}$ equation \eqref{eq:BesselK} has a pole order $-m$ in $k^2$, and at $m=0$ it has logarithmic singularity in both $t(1-t)x^2$ and $k^2$.

In particular, we find that $\cI_1(x,k_\pm(s))$ is non-singular in the holomorphic collinear limit (in which $k_\pm(s)^2 = \pm2s\la12\ra[12]\to0$). However, we have yet to take the derivatives with respect to $\lambda_2$ in equation \eqref{eq:SDGR-simplified-integral}. Doing so gives
\bea
&\la\beta\p_{\lambda_2}\ra^2\cI_1(x;k_\pm(s)) \\
&= \pi^2s^2\Bigg( - [\ut2]^2\int_0^1\dif t\,t^2(1-t)\cos(t\,x\cdot k_\pm(s))\int_0^\infty\dif r\,e^{-rt(1-t)x^2-k_\pm(s)^2/4r} \\
&- [\tilde u2]\la\beta1\ra[12]\int_0^1\dif t\,t(1-t)\sin(t\,x\cdot k_\pm(s))\int_0^\infty\frac{\dif r}{r}\,e^{-rt(1-t)x^2 - k_\pm(s)^2/4r} \\
&+ \frac{1}{4}\la\beta1\ra^2[12]^2\int_0^1\dif t\,(1-t)\cos(t\,x\cdot k_\pm(s))\int_0^\infty\frac{\dif r}{r^2}\,e^{-rt(1-t)x^2 - k_\pm(s)^2/4r}\Bigg)\,. \eea
The first of the above three terms is non-singular in the holomorphic collinear limit, but the second has a logarithmic divergence of the form $\log\la12\ra$. Logarithmic divergences of this type are expected to cancel: we demonstrate an analogous cancellation in the case of self-dual Yang-Mills explicitly in appendix \ref{app:logarithmic-cancellation}. This leaves the final term, which does contribute a first order pole
\bea
&\int_0^\infty\frac{\dif r}{r^2}\,e^{-rt(1-t)x^2-k_\pm(s)^2/4r} = 2\bigg(\frac{4t(1-t)x^2}{k_\pm(s)^2}\bigg)^{1/2}K_1(\sqrt{t(1-t)x^2k_\pm(s)^2})\\ 
&\sim \frac{4}{k_\pm(s)^2} + \cO(\log(k_\pm(s)^2)) \sim \pm\frac{2}{s\la12\ra[12]} + \cO(\log\la12\ra)\,. \eea
Therefore
\be \label{eq:SDGR-singularity-term3} \la\beta\p_{\lambda_2}\ra^2\cI_1(x;k_\pm(s))\sim \pm\frac{s\pi^2\la1\beta\ra^2[12]}{2\la12\ra}\int_0^1\dif t\,(1-t)\cos(t\,x\cdot k_{\pm}(s)) + \cO(\log\la12\ra)\,. \ee
In sum, the holomorphic collinear singularity in equation \eqref{eq:SDGR-simplified-integral} takes the form
\be - \frac{\la1\beta\ra^2[12]^3}{4\la\alpha1\ra^2\la12\ra}\int_0^1\dif s\,s\int_0^1\dif t\,(1-t)\sin(t\,x\cdot k_1)\sin(st\,x\cdot k_2) + \cO(\log\la12\ra)\,. \ee
By exploiting the identity $\la\alpha1\ra\la2\beta\ra - \la\alpha2\ra\la1\beta\ra + \la12\ra = 0$, and rescaling $s$ by a factor of $1/t$ so that it now takes values in the range $[0,t]$, we can rewrite this in a slightly more symmetric form as
\be \label{eq:SDGR-final-term} - \frac{\la1\beta\ra\la2\beta\ra[12]^3}{4\la\alpha1\ra\la\alpha2\ra\la12\ra}\int_{0\leq s\leq t\leq 1}\dif s\,\dif t\,\frac{s(1-t)}{t^2}\sin(t\,x\cdot k_1)\sin(s\,x\cdot k_2) + \cO(\log\la12\ra)\,. \ee
We show in appendix \ref{app:SDGR-calcs} that the first and second terms in equation \eqref{eq:SDGR-vertex} contribute first order poles of the form
\bea \label{eq:SDGR-remaining-terms}
- \frac{\la1\beta\ra\la2\beta\ra[12]^3}{4\la\alpha1\ra\la\alpha2\ra\la12\ra}&\int_{0\leq s\leq t\leq 1}\dif s\,\dif t\,\frac{s(1-t)^3}{t^2}\sin(t\,x\cdot k_1)\sin(s\,x\cdot k_2) + \cO(\log\la12\ra)\,, \\
\frac{2\la1\beta\ra\la2\beta\ra[12]^3}{4\la\alpha1\ra\la\alpha2\ra\la12\ra}&\int_{0\leq s\leq t\leq 1}\dif s\,\dif t\,\frac{s(1-t)^2}{t^2}\sin(t\,x\cdot k_1)\sin(s\,x\cdot k_2) + \cO(\log\la12\ra)
\eea
respectively. Putting together equations \eqref{eq:SDGR-final-term} and \eqref{eq:SDGR-remaining-terms}, and then symmetrising under the exchange $1\leftrightarrow2$ as indicated in equation \eqref{eq:SDGR-perturbiner-dif-1}, gives the simple pole
\be \label{eq:SDGR-perturbiner-singularity} - \frac{\la1\beta\ra\la2\beta\ra[12]^3}{4\la\alpha1\ra\la\alpha2\ra\la12\ra}\int_0^1\dif s\,\int_0^1\dif t\,\min(s,t)(1-\max(s,t))\sin(t\,x\cdot k_1)\sin(s\,x\cdot k_2)\,. \ee
This encodes the first order correction to the collinear singularity in the perturbiner $\cP_{\mathrm{SDGR}}(x;k_1,k_2)$ which cannot be attributed to the deformation of the zeroth order orbifold perturbiner to its curved counterpart. It therefore determines the order $c^2$ correction to the celestial OPE on Eguchi-Hanson.


\subsection{CCA for self-dual gravity on Eguchi-Hanson}

We now recast the collinear singularity in the perturbiner of self-dual gravity on the Eguchi-Hanson background, evaluated in equation~\eqref{eq:SDGR-perturbiner-singularity}, using the soft expansion. We then identify the resulting deformation with that computed using twistor methods in section \ref{sec:twistor-algebra}.

\medskip

Let's proceed by expanding \eqref{eq:SDGR-perturbiner-singularity} in powers of $\kt_1,\kt_2$. We have
\bea &\int_0^1\dif s\,\int_0^1\dif t\,\min(s,t)(1-\max(s,t))\sin(t\,x\cdot k_1)\sin(s\,x\cdot k_2) \\
&= \sum_{m,n=0}^\infty\frac{(-)^{m+n}(x\cdot k_1)^{2m+1}(x\cdot k_2)^{2n+1}}{(2m+1)!(2n+1)!}\int_0^1\dif s\,\int_0^1\dif t\,\min(s,t)(1-\max(s,t))t^{2m+1}s^{2n+1} \\
&= \sum_{m,n=0}^\infty\frac{(-)^{m+n}(x\cdot k_1)^{2m+1}(x\cdot k_2)^{2n+1}}{(2m+1)!(2n+1)!(2m+3)(2n+3)(2m+2n+5)}\,.
\eea
Writing $\mu_i^\da = x^{\da\alpha}\kappa_{i\alpha}$, in the holomorphic collinear limit $x\cdot k_1\to\tau\mu_2^\da\kt_{1\da}$, $x\cdot k_2\to\mu_2^\da\kt_{2\da}$. The above is therefore equivalent to
\be \sum_{m,n=0}^\infty\sum_{p=0}^{2m+1}\sum_{r=0}^{2n+1}\frac{(-)^{m+n}\tau^{2m+1}(\kt_{1\dzero})^p(\kt_{1\done})^{2m+1-p}(\kt_{2\dzero})^r(\kt_{2\done})^{2n+1-r}(\mu_2^\dzero)^{p+r}(\mu_2^\done)^{2(m+n+1)-p-r}}{p!(2m+1-p)!r!(2n+1-r)!(2m+3)(2n+3)(2m+2n+5)}\,. \ee
 Changing dummy variables to $q = 2m + 1 - p$, $s = 2n + 1 - r$ gives
\be \label{eq:SDGR-singularity-soft}
- \sum_{\substack{p,q=0 \\ p+q\equiv1\,(2)}}^\infty\sum_{\substack{r,s=0 \\ r+s\equiv1\,(2)}}^\infty(-)^{(p+q+r+s)/2}\frac{\tau^{p+q}(\kt_{1\dzero})^p(\kt_{1\done})^q(\kt_{2\dzero})^r(\kt_{2\done})^s(\mu_2^\dzero)^{p+r}(\mu_2^\done)^{q+s}}{p!q!r!s!(p+q+2)(r+s+2)(p+q+r+s+3)}\,.
\ee
On Eguchi-Hanson, we identify null momentum eigenstates $\delta\Theta_k(x)$ with hard generating functions in the CCA (which we denote by $W(\kt,\kappa)$ on the curved background). As discussed in subsection \ref{subsec:simplify-first-order}, we've already accounted for the change in the zeroth order OPE \eqref{eq:CCA-orbifold} from this redefinition of the states by discarding the corresponding contribution to the collinear singularity in the self-dual gravity perturbiner. Since we're working to first order in $c^2$, we can therefore continue to use the flat space identification $W(\kt,\kappa)\leftrightarrow\cos(x\cdot k)/\la\alpha\kappa\ra^4 + \cO(c^2)$. Decomposing into soft modes
\be W(\kt,\kappa) = \sum_{\substack{p,q=0 \\ p+q\equiv0\,(2)}}\frac{(-)^{(p+q)/2}(\kt_\dzero)^p(\kt_\done)^q}{p!q!}W[p,q](\kappa) \ee
this becomes $W[p,q](\kappa)\leftrightarrow (\mu^\dzero)^p(\mu^\done)^q + \cO(c^2)$. Under this identification, equation \eqref{eq:SDGR-singularity-soft} reads
\be - \la\alpha2\ra^4\sum_{\substack{p,q=0 \\ p+q\equiv1\,(2)}}^\infty\sum_{\substack{r,s=0 \\ r+s\equiv1\,(2)}}^\infty\frac{(-)^{(p+q+r+s)/2}\tau^{p+q}(\kt_{1\dzero})^p(\kt_{1\done})^q(\kt_{2\dzero})^r(\kt_{2\done})^s}{p!q!r!s!(p+q+2)(r+s+2)(p+q+r+s+3)}\,W[p+r,q+s](\kappa_2)\,. \ee
Reintroducing the coefficient from equation \eqref{eq:SDGR-perturbiner-singularity}, and absorbing the factor of $[12]^3$ into the sum through a shift of the dummy variables gives
\bea
&- \frac{2\la\al2\ra^3\la1\beta\ra\la2\beta\ra}{\la\al1\ra\la12\ra}\sum_{\substack{p,q=0 \\ p+q\equiv0\,(2)}}^\infty\sum_{\substack{r,s=0 \\ r+s\equiv0\,(2)}}^\infty\frac{(-)^{(p+q+r+s)/2}\tau^{p+q-3}(\kt_{1\dzero})^p(\kt_{1\done})^q(\kt_{2\dzero})^r(\kt_{2\done})^s}{p!q!r!s!} \\
&\qquad R_3(p,q,r,s)\,\phi_3\bigg(\frac{p+q}{2},\frac{r+s}{2}\bigg)\,W[p+r-3,q+s-3](\kappa_2)\,.
\eea
The indices in this sum are restricted to the range $p+q,r+s\geq4$, $p+r,q+s\geq 3$, and we have defined
\be \phi_3(m,n) = - \frac{3}{4(2m-1)(2n-1)(2(m+n)-3)}\,. \ee
The order $c^2$ correction to the OPE of soft modes can read off as
\bea \label{eq:SDGR-OPE-first}
&W[p,q](\kappa_1)\,W[r,s](\kappa_2) \\
&\sim - \frac{2c(\kappa_1)c(\kappa_2)}{\la12\ra}\tau^{p+q-5}R_3(p,q,r,s)
\,\phi_3\bigg(\frac{p+q}{2},\frac{r+s}{2}\bigg)\,W[p+r-3,q+s-3](\kappa_2) \\
&\sim - \frac{2c^2(\kappa_2)}{\la12\ra}\tau^{p+q-3}R_3(p,q,r,s)
\,\phi_3\bigg(\frac{p+q}{2},\frac{r+s}{2}\bigg)\,W[p+r-3,q+s-3](\kappa_2)\,, \eea
where in the second equality we've employed a Schouten. Equivalently in inhomogeneous coordinates
\bea \label{eq:SDGR-OPE-first-inhom}
&W[p,q](z_1)\,W[r,s](z_2) \\
&\sim - \frac{2z_2^2c^2}{z_{12}}R_3(p,q,r,s)\,\phi_3\bigg(\frac{p+q}{2},\frac{r+s}{2}\bigg)\,W[p+r-3,q+s-3](z_2)\,.
\eea
This is precisely the order $c^2$ contribution to the algebra we found working directly on twistor space in section \ref{sec:twistor-algebra}.

\medskip

Under the assumption that a deformed celestial chiral algebra exists, and that it remains the loop algebra of some Lie algebra deformation of $w_\wedge$, this first order computation is enough to determine it uniquely. This is because the $W(\mu)$ algebras (including the scaling limit $\mu\to\infty$) are the most general filtered deformations of $w_\wedge$ \cite{Post:1996gl}, the grading on $w_\wedge$ coinciding with the one induced by space-time dilations. Our explicit computations in this section show that we get the scaling limit.


\subsection{Self-dual Yang-Mills} \label{subsec:SDYM-space-time}

Here we provide space-time calculation of the CCA of self-dual Yang-Mills on the Eguchi-Hanson background,  again working to first order in $c^2$. The calculation is parallel to that in self-dual gravity and we leave many of the details to appendix~\ref{app:SDYM-perturbiner}. Again we show that this self-dual Yang-Mills perturbiner calculation recovers the algebra $\cL S_\wedge(\infty)$, to first order in $c^2$.

\medskip

On a background Eguchi Hanson space, self-dual Yang-Mills may be described perturbatively by the Chalmers-Siegel action~\cite{Siegel:1992wd,Chalmers:1996rq} 
\begin{equation}
\label{eq:Chalmers-Siegel-SDYM-action}
    S[\Tilde{\Phi},\Phi] = -\int \dif^4x\,\tr\left(\Tilde{\Phi}\bigg(\Delta_g\Phi - \frac{1}{2} \left[\tilde\p^\da\Phi,\tilde\p_\da\Phi\right]\bigg)\right)\,, 
\end{equation}
where again $\Delta_g$ is the Eguchi-Hanson Laplacian and the fields $\Phi,\Tilde\Phi\in \Omega^0(\cM,\fg)$. The vertex now involves the Lie bracket $[\ ,\ ]$ on $\fg$. This action is equivalent to the twistor action~\eqref{eq:sdYM-twistor-action} at the classical level. The SDYM perturbiner $\cP_{\rm SDYM}$ satisfies
\be \label{eq:SDYM-perturbiner-dif} \Delta_g\cP_\mathrm{SDYM}(x;k_1,k_2) = [\tilde\p^\da \delta\Phi_1(x),\tilde\p_\da \delta\Phi_2(x)]\,,
\ee
where the wavefunctions $\delta\Phi_i$ for fluctuations of the positive helicity gluon may be obtained by dressing the graviton states~\eqref{eq:Plebanski-fluctuations} with Lie algebra generators $t_\sfa$ and shifting their normalizations
\be 
\delta\Phi_{i\sfa}(x) = \frac{t_\sfa}{\la\al i\ra^2} \cos\sqrt{(k_i\cdot x)^2 - \frac{4c^2\la\alpha|k_i x|\beta\ra^2}{x^4}}\,. \ee

\medskip

At zeroth order in $c^2$ it's easy to compute the holomorphic collinear singularity in the perturbiner to find the celestial OPE
\be j_\sfa(\kt_1,\kappa_1)\,j_\sfb(\kt_2,\kappa_2) \sim \frac{1}{4\la12\ra}\tau^{-1}f_{\sfa\sfb}^{~~\sfc}\big(j_\sfc(\tau\kt_1 - \kt_2,\kappa_2) + j_\sfc(\tau\kt_1 + \kt_2,\kappa_2)\big)\,. \ee
Equivalently, in terms of soft modes
\be j_\sfa[p,q](\kappa_1)\,j_\sfb[r,s](\kappa_2)\sim - \frac{f_{\sfa\sfb}^{~~\sfc}}{2\la12\ra}\tau^{p+q-1}j_\sfc[p+r,q+s](\kappa_2)\,, \ee
where the indices are restricted by $p+q\equiv r+s\equiv0\Mod{2}$. Working in inhomogeneous coordinates
\be j_\sfa[p,q](z_1)\,j_\sfb[r,s](z_2)\sim - \frac{f_{\sfa\sfb}^{~~\sfc}}{2z_{12}}j_\sfc[p+r,q+s](z_2) \ee
Unsurprisingly, the CCA on the orbifold is isomorphic to $\cL\fg[\bbC^2]^{\bbZ_2}$.

\medskip

The  order $c^2$ part of equation \eqref{eq:SDYM-perturbiner-dif} is
\be \label{eq:SDYM-perturbiner-dif-1} \Delta^{(0)}\cP^{(1)}_\mathrm{SDYM}(x;k_1,k_2) = [\tilde\p^\da\delta\Phi^{(0)}_1(x),\tilde\p_\da\delta\Phi^{(1)}_2(x)] - (1\leftrightarrow2) - \Delta^{(1)}\cP^{(0)}_\mathrm{SDYM}(x;k_1,k_2)\,. 
\ee
As we saw in section \ref{subsec:simplify-first-order}, the final term is responsible for shifting the states in the zeroth order OPE to their curved counterparts. It is therefore sufficient to find the holomorphic collinear singularity induced by the first two terms. Employing the arguments sketched above for gravity, we find that the leading simple pole takes the form
\be \label{eq:SDYM-perturbiner-singularity}
\frac{\la1\beta\ra\la2\beta\ra[12]^2}{4\la12\ra}\int_0^1\dif s\,\int_0^1\dif t\,(1-\max(s,t))\cos(t\,x\cdot k_1)\cos(s\,x\cdot k_2)\,. \ee
For completeness we have included the full calculation in appendix \ref{app:SDYM-perturbiner}. Furthermore, in appendix \ref{app:logarithmic-cancellation} we show explicitly that there's no subleading logarithmic singularity of the form $\log\la12\ra$. Expanding in terms of soft modes (which we denote by $J_\sfa[m,n]$ on the curved background) the first order correction to the celestial OPE is
\bea
&J_\sfa[p,q](\kappa_1)\,J_\sfb[r,s](\kappa_2) \\
&\sim - \frac{2c(\kappa_1)c(\kappa_2)}{\la12\ra}\tau^{p+q-3}R_2(p,q,r,s)\,\psi_2\bigg(\frac{p+q}{2},\frac{r+s}{2}\bigg)\,f_{\sfa\sfb}^{~~\sfc}\,J_\sfc[p\!+\!r\!-\!2,q\!+\!s\!-\!2](\kappa_2)\,, \\
&\sim - \frac{2c^2(\kappa_2)}{\la12\ra}\tau^{p+q-1}R_2(p,q,r,s)\,\psi_2\bigg(\frac{p+q}{2},\frac{r+s}{2}\bigg)\,f_{\sfa\sfb}^{~~\sfc}\,J_\sfc[p\!+\!r\!-\!2,q\!+\!s\!-\!2](\kappa_2)\,,
\eea
or equivalently
\bea
&J_\sfa[p,q](z_1)J_\sfb[r,s](z_2) \\
&\sim - \frac{2z_2^2c^2}{z_{12}}\psi_2\bigg(\frac{p+q}{2},\frac{r+s}{2}\bigg)R_2(p,q,r,s)f_{\sfa\sfb}^{~~\sfc}J_\sfc[p+r-2,q+s-2](z_2)
\eea
in inhomogeneous coordinates. Here
\be \label{eq:psi-2} 
\psi_2(m,n) = - \frac{1}{4(2m-1)(2n-1)(2(m+n)-1)}\,,
\ee
so this agrees with the order $c^2$ correction to the $S$-algebra we found working directly on twistor space in section \ref{sec:twistor-algebra}.


\section{Switching on non-commutativity} \label{sec:non-commutativity}

So far we've seen how to obtain $\cL W(\infty)$ as the CCA of self-dual gravity on the Eguchi-Hanson background. On the other hand, the loop algebra of the symplecton, $\cL W(-3/16)$, is the CCA of Moyal-deformed self-dual gravity on the orbifold $\bbR^4/\bbZ_2$ \cite{Bu:2022iak}. It's then natural to ask whether $\cL W(\mu)$ for generic $\mu$ can arise as the CCA of a gravitational theory. In this section we argue that this is indeed the case, for Moyal deformed self-dual gravity on the Eguchi-Hanson background.


\subsection{The CCA for Moyal deformed self-dual gravity} \label{subsec:Moyal}

Let's begin by reviewing the Moyal deformation of self-dual gravity on flat space \cite{Strachan:1992em}, which is perhaps easiest to understand from the twistor perspective. See also {\it e.g.}~\cite{Kapustin:2000ek,Bridgeland:2020zjh,Gilson:2008hp,Hannabuss:2001xj,Monteiro:2022lwm,Guevara:2022qnm} for other treatments of self-dual gravity and self-dual Yang-Mills on non-commutative twistor spaces. Moyal deformed self-dual gravity involves switching on non-commutativity associated to the Poisson bracket $\p^\da\vee\p_\da/2$ on the fibres of $\PT\to\CP^1$. However, there's a catch: the bracket is twisted by $\cO(-2)$. In the twistor uplift of self-dual gravity as Poisson-BF theory this is compensated by also twisting $h$, but to turn on non-commutativity we must instead specify an unweighted bracket, which can be achieved by fixing a holomorphic section of $\cO(2)$.\footnote{Another way to achieve a consistent action is to consider infinitely many fields of increasing weight leading to chiral higher spin theories \cite{Tran:2022tft, Krasnov:2021nsq, Monteiro:2022xwq,Adamo:2022lah}.}  It's natural to make the same choice as for the defect in section \ref{sec:deforming-twistor-space}, so that the weightless Poisson structure reads
\be \pi_0 = \frac{1}{2}\la\al\lambda\ra\la\lambda\beta\ra\,\p^\da\vee\p_\da\,. \ee
We can then switch on a Moyal product associated to this Poisson structure, with formal parameter $\fq$. This parameter has weight 2 under scaling the twistor fibres. We can then write down a non-commutative analogue of Poisson BF-theory on twistor space in which the Poisson bracket in \eqref{eq:twistor-sd-gravity-action} is replaced by the Moyal bracket
\bea \label{eq:Moyal-bracket}
\{f,g\} \mapsto [f,g]_\fq &= \frac{2}{\fq(\lambda)} m\circ\sin(\fq\pi_0)(f\otimes g) \\
&= \sum_{k=0}^\infty\frac{\fq^{2k}(\lambda)}{2^{2k}\,(2k+1)!}\p_{\da_1}\dots\p_{\da_{2k+1}}f\ \p^{\da_1}\dots\p^{\da_{2k+1}}g\,. 
\eea
Here $m$ is the product map $f\otimes g\mapsto fg$, and for convenience we've defined $\fq(\lambda) = \fq\la\al\lambda\ra\la\lambda\beta\ra$. Since the Moyal bracket is defined using holomorphic bidifferential operators, it extends to $(0,q)$-forms in a straightforward way. It's also the commutator of an associative star product
\be [f,g]_\fq = \frac{1}{\fq(\lambda)}(f\star_\fq g - g\star_\fq f)\,, \ee
where
\be f\star_\fq g = m\circ\exp(\fq\pi_0)(f\otimes g)\,. \ee
We can similarly introduce a non-commutative analogue of holomorphic BF theory for the Lie algebra $\fg$ by replacing the Lie bracket appearing in \eqref{eq:sdYM-twistor-action} with its non-commutative counterpart. Since the Lie bracket is already antisymmetric, this depends on the star product through its anticommutator.

\medskip

On space-time, these correspond to Moyal deformed self-dual gravity and Yang-Mills on flat space, whose equations of motion desribe non-commutative instantons. Their celestial chiral algebras were identified in \cite{Bu:2022iak}, and can be straightforwardly recovered from the twistor description. In order to do so, let's first recall the definition of the Weyl algebra $\diff_\fq(\bbC)$. It's the quotient of the free algebra on two generators $u,v$ over $\bbC\llbracket\fq\rrbracket$ by the ideal $\mathrm{span}\{uv-vu = \fq\}$.

In the case of Moyal deformed self-dual gravity, the algebra of functions on the twistor fibre over $\lambda\neq\alpha,\beta$ inherits a Lie bracket from the interaction vertex of the non-commutative deformation of Poisson-BF theory. The resulting Lie algebra is isomorphic to $\diff_\fq(\bbC)$ equipped with the standard Moyal bracket. Forming the loop algebra gives the CCA.

For Moyal deformed self-dual Yang-Mills, the algebra of functions on the twistor fibre over $\lambda\neq\alpha,\beta$ inherits a Lie bracket from the interaction vertex of the non-commutative BF theory. The resulting Lie algebra structure on $\diff_\fq(\bbC)\otimes\fgl(N)$ is defined using the star product on the first factor and matrix multiplication on the second. The CCA is then obtained by taking the loop algebra.

Quotienting space-time by $\bbZ_2$, the Weyl algebra $\diff_\fq(\bbC)$ is restricted to its $\bbZ_2$-invariant subspace, the symplecton $W(-3/16;\fq)$ \cite{Pope:1989sr}. Making this replacement in the CCAs of Moyal deformed self-dual gravity and Yang-Mills on flat space gives their counterparts on the orbifold $\bbR^4/\bbZ_2$.


\subsection{The CCA for Moyal deformed self-dual gravity on Eguchi-Hanson} \label{subsec:NCEH}

Now let's consider the result of coupling the non-commutative analogue of Poisson-BF theory with Poisson bracket replaced by \eqref{eq:Moyal-bracket} to a holomorphic surface defect as in equation \eqref{eq:twistor-defect-action}. In the presence of this defect the equation of motion for $h$ becomes
\be \label{eq:sourced-NCBF}
\bar\p h + \frac{1}{2}[h,h]_\fq = 4\pi^2 c^2(\lambda)\,\bar\delta^2(\mu)\,.
\ee
We can treat the parameter $c$ (which has weight 2 under scaling the twistor fibres) as a finite complex parameter, or as a formal parameter proportional to $\fq$ with constant of proportionality $\tilde c$. In both cases the resulting CCA will be defined over $\bbC\llbracket\fq\rrbracket$. We solve for $h$ exactly as we did in section \ref{sec:deforming-twistor-space} to get
\be h = c^2(\lambda)\,\frac{[\hat\mu\,\dif\hat\mu]}{2[\mu\,\hat\mu]^2}\,. \ee
This induces a non-commutative analogue of a Dolbeault operator
\be \nbar_\fq = \dbar + [h,\ ]_\fq = \dbar - c^2(\lambda)[\hat\mu\,\dif\hat\mu]\sum_{k=0}^\infty\frac{(k+1)\fq^{2k}(\lambda)}{[\mu\,\hat\mu]^{2k+3}2^{2k}}\hat\mu^{\al_1}\dots\hat\mu^{\dot\al_{2k+1}}\p_{\da_1}\dots\p_{\dot\al_{2k+1}}\,. \ee
Functions, and more generally differential forms, in the kernel of this operator should be viewed as `holomorphic'. Since $\nbar_\fq$ distributes over the star product, star products of holomorphic functions in this non-commutative sense are themselves holomorphic. In particular, it's easy to check that the functions
\be X^{\da\db} = \mu^\da\mu^\db - c^2(\lambda)\frac{\hat\mu^\da\hat\mu^\db}{[\mu\,\hat\mu]^2} \ee
from equation \eqref{eq:newholomorphic} remain holomorphic. We can then determine
\bea \label{eq:deformed-star-product}
&X^{\da\db}\star_\fq X^{\dc\dd} = X^{\da\db}X^{\dc\dd} + \frac{\fq(\lambda)}{2}\big(\eps^{\da\dc}X^{\db\dd} + \eps^{\da\dd}X^{\db\dc} + \eps^{\db\dc}X^{\da\dd} + \eps^{\db\dd}X^{\da\dc}\big) \\
&+ \frac{\fq^2(\lambda)}{4}\bigg(\eps^{\da\dc}\eps^{\db\dd} + \eps^{\da\dd}\eps^{\db\dc} - \frac{12c^2(\lambda)\hat\mu^\da\hat\mu^\db\hat\mu^\dc\hat\mu^\dd}{[\mu\,\hat\mu]^4} \bigg)\,.
\eea
We remark that the explicit $\hat\mu$ dependence in the final term, which may seem surprising, is needed to compensate the fact that the commutative product $X^{\da\db}X^{\dc\dd}$ is not in the kernel of $\nbar_\fq$. 

From the above we infer that
\be \label{eq:commutator} [X^{\da\db},X^{\dc\dd}]_\fq = \eps^{\da\dc}X^{\db\dd} + \eps^{\da\dd}X^{\db\dc} + \eps^{\db\dc}X^{\da\dd} + \eps^{\db\dd}X^{\da\dc}\,. \ee
If there were no further constraints on the products of the $X^{\da\db}$, we'd learn that on each generic twistor fibre ($\lambda\neq\alpha,\beta$), under the star product $\star_\fq$ they generate the universal enveloping algebra (UEA) of\footnote{Here $\fg_{\fq(\lambda)}$ denotes the Lie algebra over $\bbC\llbracket\fq\rrbracket$ obtained by multiplying the structure constants of $\fg$ by $\fq(\lambda)$.} $\fsl_{2,\fq(\lambda)}$.  However, contracting indices in~\eqref{eq:deformed-star-product} gives
\be \label{eq:Casimir} 
X^{\da\db}\star_\fq X_{\da\db} = X^{\da\db}X_{\da\db} + \frac{3\fq^2(\lambda)}{2} = -2c^2(\lambda) + \frac{3\fq^2(\lambda)}{2}\,, 
\ee
where the second equality uses the constraint $X^{\da\db}X_{\da\db} = -2c^2(\lambda)$. At this point we notice that in the $\fq\to0$ limit we recover the commutative algebra generated by the  $X^{\da\db}$ subject to the constraint $X^{\da\db}X_{\da\db} = -2c^2(\lambda)$, equipped with the standard Poisson structure \eqref{eq:Poisson-bracket}. This is isomorphic to $W(\infty)$, as we saw in section \ref{sec:twistor-algebra}.

The standard normalization of the Casimir in the UEA of $\fsl_{2,\fq(\lambda)}$ is $C = -X^{\da\db}\star_\fq X_{\da\db}/8$, which has eigenvalues $\fq^2(\lambda)\mu = \fq^2(\lambda)\sigma(\sigma+1)$. Taking $c$ to be a formal parameter proportional to $\fq$, {\it i.e.}, $c = \fq\tilde c$, we have
\be C = -\frac{1}{8}X^{\da\db}\star_\fq X_{\da\db} = \frac{\fq^2(\lambda)}{16}(4\tilde c^2 - 3) = \fq^2(\lambda)\mu\,, \ee
so that $\mu = (4\tilde c^2 - 3)/16$. On each twistor fibre the $X^{\da\db}$ then generate $U(\fsl_{2,\fq(\lambda)})/\mathrm{span}\{C - \fq^2(\lambda)\mu\}$. The Lie algebra defined through the commutator is isomorphic to $W(\mu;\fq(\lambda))\cong W(\mu;\fq)$ \cite{Pope:1989sr}. Forming the loop algebra gives the CCA of Moyal deformed self-dual gravity on Eguchi-Hanson.

Setting the Eguchi-Hanson parameter $c=0$, the non-commutative algebra generated by the $X^{\da\db}$ equipped with the bracket \eqref{eq:commutator} is isomorphic to $W(-3/16;\fq)$ \cite{Pope:1989sr}, consistent with the results of \cite{Bu:2022iak}. Indeed, equation \eqref{eq:Casimir} is the unique consistent relation with the correct $\fq\to0$ limit and compatible with the grading induced by scaling the twistor fibres. Notice that the shift $3\fq^2(\lambda)/4$ to the Casimir for $X^{\da\db}\star_\fq X_{\da\db}$ compared to that for the commutative $X^{\da\db}X_{\da\db}$ is essential to find the symplecton algebra at $\mu=-3/16$. While this is the expected value~\cite{Pope:1991ig}, it is rather unusual -- 
for example, quantizing coadjoint orbits of $\fsl_2$ via the A-model or Duflo-Kirillov-Kontsevich map induces a shift to $\mu = -1/4$ \cite{Gukov:2008ve,Chervov:2004tw} -- and it is gratifying to obtain it so directly.

By varying $\tilde c$ we sweep out all possible choices of the parameter $\mu$. Two cases are of particular interest: setting $\tilde c^2 = -1/4$, we find that $W(\mu;\fq)$ is isomorphic to the wedge subalgebra of $W_{1+\infty}$, whilst at $\tilde c^2=3/4$ we have instead the wedge subalgebra of $W_\infty$. In particular, this gives a bulk interpretation for the deformation to $W_{1+\infty}$ as speculated in \cite{Strominger:2021lvk}. In Euclidean signature it's natural to take $\tilde c^2\geq0$ so that only those $W(\mu;\fq)$ algebras with $\mu\geq-3/16$ are attainable, excluding the wedge subalgebra of $W_{1+\infty}$. We do not see any restrictions on the sign of $\tilde c^2$ in ultrahyperbolic signature.


\section{Discussion} \label{sec:discussion}

One motivation for this work was to clarify the relationship between the vertex algebras arising in the celestial holography literature, and the infinite dimensional Lie algebras arising as wedge subalgebras of infinite $W$-algebras. CCAs for the class of self-dual theories considered here are loop algebras of infinite dimensional Lie algebras. Let's briefly summarise those which feature in this note.

For self-dual gravity on the Eguchi-Hanson background, we've found that the appropriate infinite dimensional Lie algebra is $W(\infty)$. Taking the limit of the Eguchi-Hanson parameter $c\to0$, the geometry degenerates to the orbifold $\bbR^4/\bbZ_2$. This is reflected in the CCA: the $W(\infty)$ algebra contracts to $w_\wedge$, the wedge subalgebra of $w_{1+\infty}$. $w_\wedge$ is also a $\bbZ_2$ quotient of $\ham(\bbC^2)$, corresponding to the CCA of self-dual gravity on flat space-time.

Allowing the twistor space to become non-commutative with formal parameter $\fq$, corresponds to considering the Moyal deformation of self-dual gravity. Working on an Eguchi-Hanson background with formal parameter $c = \tilde c\fq$, the appropriate infinite dimensional Lie algebra is likewise deformed to $W(\mu;\fq)$, where
\be \label{eq:mu-again} \mu = \frac{4\tilde c^2 - 3}{16}\,. \ee
In particular, when $\tilde c=0$ we recover the symplecton, itself the $\bbZ_2$ quotient of the Weyl algebra. This is the infinite dimensional algebra determining the CCA of Moyal deformed self-dual gravity on flat space-time \cite{Bu:2022iak}. 

All of the above statements carry over to self-dual Yang-Mills on Eguchi-Hanson, and, taking appropriate care to track powers of $\fq$, to its Moyal deformation.

\medskip

We conclude with a brief discussion of some future directions suggested by the results of this paper.

Firstly, as mentioned in the introduction, it's clear that many of the considerations of this paper can be extended to the case of Gibbons-Hawking multi-centre metrics, and (self-dual) gravitational instantons more generally. We will address these points in upcoming work \cite{Bittleston:2023soon}.

For particular values of the parameter $\tilde c$, the algebra $W(\mu,\fq)$ with $\mu$ as in \eqref{eq:mu-again} specialises to the wedge subalgebras of $W_\infty$ or $W_{1+\infty}$. These extend outside the wedge without the need to introduce infinitely many fields of negative conformal spin, and can realised using families of composite operators in free 2d holomorphic theories. Unfortunately, the `space-time' of these 2d theories is the right- rather than left-handed celestial sphere. 

On the other hand, these algebras also naturally arise as operator algebras on the Higgs and Coulomb branches of 3d $\cN=4$ and 4d $\cN=2$ SCFTs \cite{Beem:2016cbd,Costello:2017fbo,Bullimore:2015lsa}.

In \cite{Mago:2021wje,Ren:2022sws} the authors explore certain Lie algebra deformations of tree level CCAs for theories of self-dual gravity and Yang-Mills non-minimally coupled to matter. The Jacobi identity constrains the deformed OPEs, and hence the possible coupled theories admitting CCAs. The Lie algebras obtained there are not directly related to wedge subalgebras of $W_{1+\infty}$ and $W_{\infty}$, which only arise as deformations of the $\bbZ_2$ fixed point subalgebra of $\ham(\bbC^2)$ as discussed above. They're more closely related to higher spin analogues of the symplecton algebra restricted to $|s|\leq2$ \cite{Monteiro:2022xwq}.

The CCAs in this note are all loop algebras of infinite dimensional Lie algebras. It's natural to ask whether still more complicated celestial chiral algebras can arise geometrically; say, with higher order poles or non-linear OPEs? A potentially related issue is that self-dual Einstein gravity has trivial (or more precisely distributional) tree amplitudes, a fact which remains true on any on-shell background.\footnote{As suggested to us by Kevin Costello: since the twistor space of any self-dual Ricci-flat metric fibres over $\CP^1$, scattering states lift to twistor space with support in an arbitrarily small neighbourhood of their left-handed spinor helicity variable. Any tree diagram must include a vertex with two external legs; therefore, beyond $n=3$ it necessarily vanishes for generic kinematics, just as in flat space.}

\begin{figure}[!ht]
	\centering
 	\includegraphics[scale=0.33]{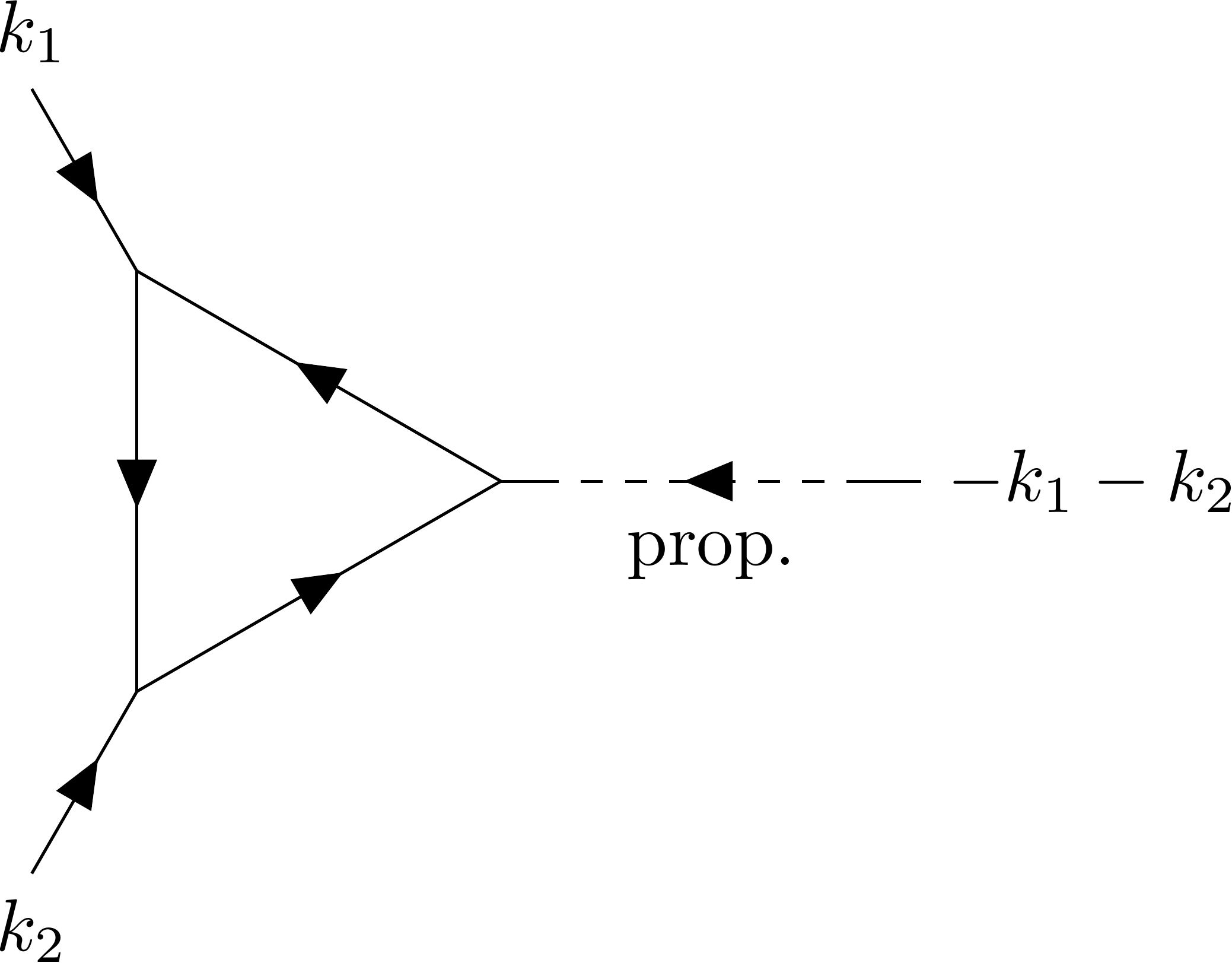}
	\caption{\emph{1-loop diagram leading to double poles in a graviton amplitude as the momenta $k_1,k_2$ of two positive helicity external states become holomorphically collinear.}} \label{fig:loopsplitting}
\end{figure}

One concrete way to obtain higher-order poles and non-linearities  is by incorporating loop corrections \cite{Costello:2022upu,Bittleston:2022jeq}, though this requires introducing states in the chiral algebra corresponding to negative helicity fields. Classically these transform in the adjoint representation of algebra generated by the positive helicity states. At 1-loop the perturbiner gets corrected by the diagram illustrated in figure \ref{fig:loopsplitting}. Since all the external legs are incoming, this describes a correction to the OPE of positive helicity states which is proportional to the negative helicity states.

Deforming self-dual gravity at first-order, {\it e.g.}, to full Einstein gravity, we get non-vanishing loop amplitudes whose collinear singularities receive 1-loop corrections from this diagram. However, the collinear singularities of these amplitudes will not be universal unless the 1-loop all-plus amplitudes vanish. This is reflected in the chiral algebra, which does not have an associative operator product unless certain anomalies on twistor space, which can be identified with the space-time 1-loop all-plus amplitudes, vanish.

In \cite{Bittleston:2022nfr} a number of methods of cancelling the twistorial anomaly/1-loop all-plus amplitudes in self-dual gravity were presented, inspired by analogous methods for self-dual Yang-Mills \cite{Costello:2021bah}. The simplest anomaly free variant is $\cN=1$ self-dual supergravity. Another possibility is to couple to a gravitational axion with 4\textsuperscript{th}-order kinetic term, which cancels the 1-loop amplitudes through tree level exchange. Putting either of these theories on twistor space will lead to consistent quantum deformed CCAs.


\acknowledgments

It is a pleasure to thank Kevin Costello, Nick Dorey, Maciej Dunajski, Sean Seet, Atul Sharma, Andy Strominger and Peter Wildemann for helpful conversations, and Atul Sharma for collaboration at an early stage of this project. We also thank the organisers of the workshop `Celestial Amplitudes and Flat Space Holography' at the Corfu Summer Institute in 2022, and the conference `Quantum de Sitter Universe' at the University of Cambridge in 2023. SH also thanks the organisers of the '2023 Winter School in Mathematical Physics' at the SwissMAP research station and particularly Tudor Dimofte for his interesting lectures and helpful conversations.

\section*{Declarations}

Research of RB at Perimeter Institute is supported in part by the Government of Canada through the Department of Innovation, Science and Economic Development and by the Province of Ontario through the Ministry of Colleges and Universities.  The work of SH \& DS has been supported in part by STFC HEP Theory Consolidated grant ST/T000694/1. SH is partly supported by St. John’s College, Cambridge. The authors have no competing interests to declare that are relevant to the content of this article. Data sharing is not applicable to this article as no datasets were generated or analysed during the current study. 

\newpage

\begin{appendix}


\section{Isomorphic celestial chiral algebras} \label{app:isomorphism}

In this appendix, we will discuss two isomorphisms from the  (loop algebras of the) algebras in the natural basis on twistor space --  \eqref{eqs:algebra-in-twistor-basis} for SDGR and \eqref{eq:twistor-S-structure-constants} for SDYM -- to the algebras found
in the scattering basis -- \eqref{eq:sl-infinity} for SDGR and \eqref{eq:SDYM-CCA-twistor} for SDYM.

\medskip

Recall that the polynomials $V[2p,2q]$ and $V[2p+1,2q+1]$ only appear in certain combinations in the scattering states. From their definition in equation \eqref{eq:softbasis}, we see that 
\bea
\label{eq:extractingSoftModes}
W[2p,2q]=&\frac{(2p)!(2q)!}{(2p+2q)!}\sum_{a=0}^{\min(p,q)}\binom{p+q}{p-a,q-a,2a}X^{p-a}Y^{q-a}(2Z)^{2a}\\
=&\frac{(2p)!(2q)!}{(2p+2q)!}\sum_{a=0}^{\min(p,q)}2^{2a} \binom{p+q}{p-a,q-a,2a}  \sum_{\ell=0}^a \binom{a}{\ell} c(\lambda)^{2\ell} X^{p-\ell}Y^{q-\ell}\\
=&\sum_{\ell=0}^{\min(p,q)}  (2c(\lambda))^{2\ell} \left(\frac{(2p)!(2q)!}{(2p+2q)!} \sum_{a=\ell}^{\min(p,q)}2^{2(a-l)} \binom{a}{\ell} \binom{p+q}{p-a,q-a,2a}\right)\\
&V[2(p-\ell),2(q-\ell)]\,,
\eea
where the sum over $a$ can be performed as
\be
\sum_{a=\ell}^{\min(p,q)}2^{2(a-l)} \binom{a}{\ell} \binom{p+q}{p-a,q-a,2a}=\frac{(2p+2q)!}{(2p)!(2q)!} \frac{[p]_\ell\,[q]_\ell\,[p+q]_\ell}{\ell!\,[2(p+q)]_{2\ell}}\,,
\ee
to give
\be
C_0(p,q,\ell)=\frac{[p]_\ell\,[q]_\ell\,[p+q]_\ell}{\ell!\,[2(p+q)]_{2\ell}}\,
\ee
from equation \eqref{eq:c0c1}. For the odd case, we similarly have
\bea
\label{eq:extractingSoftModesOdd}
W[2p+1,2q+1]=&\frac{(2p+1)!(2q+1)!}{(2p+2q+2)!}\sum_{a=0}^{\min(p,q)}\binom{p+q+1}{p-a,q-a,2a+1}X^{p-a}Y^{q-a}(2Z)^{2a+1}\\
=&\frac{(2p+1)!(2q+1)!}{(2p+2q+2)!}\sum_{a=0}^{\min(p,q)}2^{2a+1} \binom{p+q+1}{p-a,q-a,2a}  \sum_{\ell=0}^a \binom{a}{\ell} c(\lambda)^{2\ell} X^{p-\ell}Y^{q-\ell}Z\\
=&\sum_{\ell=0}^{\min(p,q)}  (2c(\lambda))^{2\ell} \left(\frac{(2p+1)!(2q+1)!}{(2p+2q+2)!}\sum_{a=\ell}^{\min(p,q)}2^{2(a-l)+1} \binom{a}{\ell} \binom{p+q+1}{p-a,q-a,2a}\right)\\
&V[2(p-\ell)+1,2(q-\ell)+1]\,,
\eea
where once again the sum over $a$ can be performed as
\be
\sum_{a=\ell}^{\min(p,q)}2^{2(a-l)+1} \binom{a}{\ell} \binom{p+q+1}{p-a,q-a,2a}= \frac{(2p+2q+2)!}{(2p+1)!(2q+1)!} \frac{[p]_{\ell}
\,[q]_{\ell}\,[p+q+1]_{\ell}}{\ell!\,[2(p+q+1)]_{2\ell}}\,,
\ee
to give
\be
C_1(p,q,\ell)= \frac{[p]_{\ell}
\,[q]_{\ell}\,[p+q+1]_{\ell}}{\ell!\,[2(p+q+1)]_{2\ell}}\,.
\ee

These arguments go through line by line to give the form of the soft gluon modes in terms of the polynomials \eqref{eq:GaugePolynomials}
\bea
J_\sfa[2p,2q]&=\sum_{\ell=0}^{\min(p,q)}(2c(\lambda))^{2\ell}\, C_0(p,q,\ell)\, j_\sfa[2(p-\ell),2(q-\ell)]\,,\\
J_\sfa[2p\!+\!1,2q\!+\!1]&= \sum_{\ell=0}^{\min(p,q)}(2c(\lambda))^{2\ell}\,C_1(p,q,\ell)\,j_\sfa[2(p-\ell)\!+\!1,2(q-\ell)\!+\!1]\,.
\eea

To see that the change of basis is actually an isomorphism Lie algebras we expand both sides of 
\bea
\label{eq:appendixW}
&\big[W[p,q],W[r,s]\big] \\
&= \frac{1}{2}\sum_{\ell\geq0} (2c(\lambda))^{2\ell}R_{2\ell+1}(p,q,r,s)\psi_{2\ell+1}\bigg(\frac{p+q}{2},\frac{r+s}{2}\bigg)W[p\!+\!r\!-\!2\ell\!-\!1,q\!+\!s\!-\!2\ell\!-\!1]\,
\eea
in terms of the $V$ basis through equations \eqref{eq:extractingSoftModes} and \eqref{eq:extractingSoftModesOdd}. For, say, two even elements, the left hand side reads 
\bea
\label{eq:IsomLHS}
&[W[2p,2q],W[2r,2s]]=\sum_{\ell=0}^{\min(p,q)+\min(r,s)} (2c(\lambda))^{2\ell}  \, V[2(p\!+\!r\!-\!\ell)\!-\!1,2(q\!+\!s\!-\!\ell)]\\
&\left(\sum_{i=0}^\ell R_1\big(2(p-i),2(q-i),2(r-(\ell-i)),2(s-(\ell-i))\big) 
C_0(p,q,i)C_0(r,s,\ell-i) \right)\,,
\eea
while the right hand side reads
\bea
\label{eq:IsomRHS}
&\sum_{\ell=0}^{\min(p+r,q+s)} (2c(\lambda))^{2\ell}V[2(p\!+\!r\!-\!\ell)\!-\!1,2(q\!+\!s\!-\!\ell)\!-\!1]\\
&\left(\sum_{i=0}^\ell R_{2i+1}(2p,2q,2r,2s)\psi_{2i+1}(2(p+q),2(r+s))C_1(p\!+\!r\!-\!i\!-\!1,q\!+\!s\!-\!i\!-\!1,l\!-\!i) \right)\,.
\eea
While the individual summands in the last lines of \eqref{eq:IsomLHS} and \eqref{eq:IsomRHS} do not match, the whole sum does. This has been verified numerically up to $\ell=8$. Similarly, we have numerically verified up to $\ell=8$ that \eqref{eq:appendixW} also holds for $[W[2p+1,2q+1],W[2r,2s]]$ and for $[W[2p+1,2q+1],W[2r+1,2s+1]]$. Similar checks have also been performed in the case of self-dual Yang-Mills theory.


\section{Space-time calculations} \label{app:space-time-calcs}

In these appendices we include derivations of the results used in section \ref{sec:space-time-algebra}.


\subsection{Self-dual gravity perturbiner} \label{app:SDGR-calcs}

In this appendix we evaluate the contributions of the first and second terms on the right hand side of equation \eqref{eq:SDGR-vertex} to the leading holomorphic collinear singularity in equation \eqref{eq:SDGR-singularity-integral}. These are listed in equation \eqref{eq:SDGR-remaining-terms}. We employ the method outlined in section \ref{subsec:simplify-first-order}, and shall adopt the same notation.

The contribution of the first term in \eqref{eq:SDGR-vertex} to the integral \eqref{eq:SDGR-singularity-integral} is
\be \label{eq:SDGR-term1} -\frac{2}{\pi^2\la\al1\ra^2\la\al2\ra^2}\int_0^1\dif s\,\int_{\bbR^4}\frac{\dif^4y}{(x-y)^2y^6}\,[\vt1][v2]\bigg([12]-\frac{6[v1][\vt2]}{y^2}\bigg)\cos(y\cdot k_1)\cos(s\,y\cdot k_2)\,. \ee
Replacing the factors of $[vi],[\vt i]$ by derivatives with respect to helicity variables, this can written as
\bea \label{eq:SDGR-simplified-term1}
&\frac{1}{\pi^2\la\alpha1\ra^2\la\alpha2\ra^2}\int_0^1\frac{\dif s}{s^2}\,\Big(s[12]\la\al\p_{\lambda_2}\ra\la
\beta\p_{\lambda_1}\ra\big(\cI_2(x,k_-(s)) - \cI_2(x,k_+(s))\big) \\
&+ 6\la\al\p_{\lambda_1}\ra\la\al\p_{\lambda_2}\ra\la
\beta\p_{\lambda_1}\ra\la
\beta\p_{\lambda_2}\ra\big(\cI_3(x,k_-(s)) + \cI_3(x,k_+(s)\big)\Big)\,.
\eea
Differentiating $\cI_k(x,k_\pm(s))$ with respect to spinor helicity variables a total of $k+l$ times generates holomorphic collinear singularities of at worst order $l$. The logarithmic singularities are expected to cancel as they do in self-dual Yang-Mills, so we'll concentrate on the simple pole generated by the second set of terms. Furthermore, recalling that
\be \cI_m(x;k_\pm(s)) = \frac{\pi^2}{k!}\int_0^1\dif t\,(1-t)^m\cos(t\,x\cdot k_\pm(s))\int_0^\infty\dif r\,r^{m-1}e^{-rt(1-t)x^2-k_\pm(s)^2/4r}\,, \ee
the pole of order $l$ is only generated if all $l+m$ derivatives with respect to spinor helicity variables hit the exponential $\exp(-k_\pm(s)^2/4r)$ to bring down a factor of $1/r^{l+m}$. Therefore the singularity in \eqref{eq:SDGR-simplified-term1} is determined by
\bea
&6\la\al\p_{\lambda_1}\ra\la\al\p_{\lambda_2}\ra\la
\beta\p_{\lambda_1}\ra\la
\beta\p_{\lambda_2}\ra\cI_3(x,k_\pm(s)) \\
&\sim \frac{\pi^2s^4\la\al1\ra\la\al2\ra\la1\beta\ra\la2\beta\ra[12]^4}{16}\int_0^1\dif t\,(1-t)^3\cos(t\,x\cdot k_\pm(s))\int_0^\infty\frac{\dif r}{r^2}\,e^{-rt(1-t)x^2-k_\pm(s)^2/4r} \\
&+ \cO(\log\la12\ra) \\
&\sim \pm\frac{\pi^2s^3\la\al1\ra\la\al2\ra\la1\beta\ra\la2\beta\ra[12]^3}{16\la12\ra}\int_0^1\dif t\,(1-t)^3\cos(t\,x\cdot k_\pm(s)) + \cO(\log\la12\ra)\,.
\eea
Hence, in the holomorphic collinear limit equation \eqref{eq:SDGR-term1} has a leading simple pole
\be - \frac{\la1\beta\ra\la2\beta\ra[12]^3}{4\la\al1\ra\la\al2\ra\la12\ra}\int_0^1\dif s\,s\int_0^1\dif t\,(1-t)^3\sin(s\,x\cdot k_1)\sin(st\,x\cdot k_2)\,. \ee
Rescaling $s$ by a factor of $1/t$, so that it now takes values in the range $[0,t]$, gives the first line of equation \eqref{eq:SDGR-remaining-terms}.

Let's move on to the second term in \eqref{eq:SDGR-vertex}, whose contribution to \eqref{eq:SDGR-singularity-integral} is
\bea \label{eq:SDGR-term2}
&-\frac{[12]}{\pi^2\la\al1\ra^2\la\al2\ra}\int_0^1\dif s\,s\int_{\bbR^4}\frac{\dif^4y}{(x-y)^2y^4}\,[\vt2] \\
&\bigg([12]-\frac{2([v1][\vt2] + [\vt1][v2])}{y^2}\bigg)\cos(y\cdot k_1)\sin(s\,y\cdot k_2)\,,
\eea
or equivalently
\bea \label{eq:SDGR-simplified-term2}
&-\frac{[12]}{2\pi^2\la\alpha1\ra^2\la\alpha2\ra}\int_0^1\frac{\dif s}{s}\,\la\beta\p_{\lambda_2}\ra\Big(s[12]\big(\cI_1(x;k_-(s)) + \cI_1(x;k_+(s))\big) \\
&+ 2(\la\al\p_{\lambda_1}\ra\la\beta\p_{\lambda_2}\ra + \la\beta\p_{\lambda_1}\ra\la\al\p_{\lambda_2}\ra)\big(\cI_2(x;k_-(s)) - \cI_2(x;k_+(s))\big)\Big)\,.
\eea
Again, the logarithmic singularities should cancel. The only potential pole is therefore generated by the second set of terms. We have
\bea \label{eq:SDGR-singularity-term2}
&2\la\al\p_{\lambda_2}\ra\la\beta\p_{\lambda_1}\ra\la\beta\p_{\lambda_2}\ra\cI_2(x;k_\pm(s)) \\
&\sim\mp\frac{\pi^2s^3\la\alpha1\ra\la1\beta\ra\la2\beta\ra[12]^3}{8}\int_0^1\dif t\,(1-t)^2\cos(s\,x\cdot k_\pm(s))\int_0^\infty\frac{\dif r}{r^2}\,e^{-rt(1-t)x^2-k_\pm(s)^2/4r} \\
&+ \cO(\log\la12\ra) \\
&\sim - \frac{\pi^2s^2\la\alpha1\ra\la1\beta\ra\la2\beta\ra[12]^2}{4\la12\ra}\int_0^1\dif t\,(1-t)^2\cos(t\,x\cdot k_\pm(s)) + \cO(\log\la12\ra)\,.
\eea
Similarly
\bea
&2\la\al\p_{\lambda_1}\ra\la\beta\p_{\lambda_2}\ra^2\cI_2(x;k_\pm(s)) \\
&\sim - \frac{\pi^2s^2\la\alpha2\ra\la1\beta\ra^2[12]^2}{4\la12\ra}\int_0^1\dif t\,(1-t)^2\cos(t\,x\cdot k_\pm(s)) + \cO(\log\la12\ra)\,.
\eea
Invoking the Schouten identity $\la\alpha2\ra\la1\beta\ra = \la\alpha1\ra\la2\beta\ra + \la12\ra$ and discarding the non-singular piece we find that this simple pole coincides with that in \eqref{eq:SDGR-singularity-term2}. Hence, in the holomorphic collinear limit equation \eqref{eq:SDGR-term2} has a leading simple pole
\be \frac{\la1\beta\ra\la2\beta\ra[12]^3}{2\la\alpha1\ra\la\alpha2\ra\la12\ra}\int_0^1\dif s\,s\int_0^1\dif t\,(1-t)^2\sin(s\,x\cdot k_1)\sin(st\,x\cdot k_2)\,. \ee
Rescaling $s$ by $1/t$ gives the second line  of equation \eqref{eq:SDGR-remaining-terms} in the main text.


\subsection{Self-dual Yang-Mills perturbiner} \label{app:SDYM-perturbiner}

In this appendix we compute the leading holomorphic collinear singularity in the first order correction to the self-dual Yang-Mills perturbiner at first order in $c^2$, as given in equation \eqref{eq:SDYM-perturbiner-singularity}. We follow the approach used for self-dual gravity, as presented in section \ref{subsec:simplify-first-order} and appendix \ref{app:SDGR-calcs}.

It's sufficient to determine the leading holomorphic collinear singularity in
\be \label{eq:SDYM-singularity-integral} - \frac{1}{4\pi^2}\int_{\bbR^4}\frac{\dif^4y}{(x-y)^2}\,\left[\tilde\p^\da\Phi_{1\sfa}^{(0)}(y),\tilde\p_\da\Phi_{2\sfb}^{(1)}(y)\right]\,. 
\ee
Recalling the first order correction to a null momentum eigenstate on Eguchi-Hanson \eqref{eq:first-order-eigenstate}, we find 
\bea \label{eq:SDYM-vertex}
\left[\tilde\p^\da\Phi_{1\sfa}^{(0)}(x)\,\tilde\p_\da\Phi_{2\sfb}^{(1)}(x)\right] &= -  f^{~~\sfc}_{\sfa\sfb}\,t_\sfc\,\frac{4[\ut2]\sin(x\cdot k_1)}{\la\al1\ra}\bigg(\frac{2[\ut 1][u2]}{x^6}\int_0^1\dif s\,\cos(s\,x\cdot k_2) \\
&\qquad\qquad\qquad + \frac{\la\al2\ra[\ut2][12]}{2x^4}\int_0^1\dif s\,s\sin(s\,x\cdot k_2)\bigg)\,.
\eea
Let's address the two terms on the right hand side separately. Stripping the colour factor, the contribution of the first term to equation \eqref{eq:SDYM-singularity-integral} is
\be \label{eq:SDYM-term1} \frac{2}{\pi^2\la\al1\ra}\int_0^1\dif s\,\int_{\bbR^4}\frac{\dif^4y}{(x-y)^2y^6}\,[\vt1][v2][\vt2]\sin(x\cdot k_1)\cos(s\,x\cdot k_2)\,. \ee
Employing familiar tricks this can be written as
\be \label{eq:SDYM-simplified-term1} \frac{1}{\pi^2\la\al1\ra}\int_0^1\frac{\dif s}{s^2}\,\la\al\p_{\lambda_2}\ra\la\beta\p_{\lambda_1}\ra\la\beta\p_{\lambda_2}\ra\big(\cI_2(x;k_-(s)) + \cI_2(x;k_+(s))\big)\,. \ee
From the discussion in appendix \ref{app:SDGR-calcs}, we know that hitting $\cI_2(x;k_\pm(s))$ with 3 derivatives with respect to spinor helicity variables can generate at worst a simple pole. For now we ignore the subleading logarithmic singularities of the form $\log\la12\ra$, though we'll see that they cancel explicitly in appendix \ref{app:logarithmic-cancellation}. Using equation \eqref{eq:SDGR-singularity-term2}, the leading simple pole in equation \eqref{eq:SDYM-term1} is
\be \label{eq:SDYM-term1-final} - \frac{\la1\beta\ra\la2\beta\ra[12]^2}{4\la12\ra}\int_0^1\dif s\,\int_0^1\dif t\,(1-t)^2\cos(t\,x\cdot k_1)\cos(st\,x\cdot k_2)\,.
\ee

The contribution of the second term in equation \eqref{eq:SDYM-vertex} is
\be \frac{\la\al2\ra[12]}{2\pi^2\la\al1\ra}\int_0^1\dif s\,s\int_{\bbR^4}\frac{\dif^4y}{(x-y)^2y^4}\,[\vt2]^2\sin(x\cdot k_1)\sin(s\,x\cdot k_2)\,. \ee
The usual tricks turn this into
\be \label{eq:SDYM-simplified-term2} \frac{\la\alpha2\ra[12]}{4\pi^2\la\alpha1\ra}\int_0^1\frac{\dif s}{s}\la\beta\p_{\lambda_2}\ra^2\big(\cI_1(x;k_-(s))-\cI_1(x;k_+(s))\big)\,, \ee
which has at worst a simple pole. Ignoring the subleading logarithmic singularity, the leading simple pole was determined in equation \eqref{eq:SDGR-singularity-term3}, giving
\be \label{eq:SDYM-term2-final} \frac{\la\al2\ra\la1\beta\ra^2[12]^2}{4\la\al1\ra\la12\ra}\int_0^1\dif s\,\int_0^1\dif t\,(1-t)\cos(t\,x\cdot k_1)\cos(st\,x\cdot k_2)\,. \ee
The coefficient can be symmetrised using a Schouten.

Combining equations \eqref{eq:SDYM-term1-final} and \eqref{eq:SDYM-term2-final} (after symmetrising the coefficient), and then rescaling $s$ by $1/t$ so that it now takes values in the range $[0,t]$, we arrive at
\be \frac{\la1\beta\ra\la2\beta\ra[12]^2}{4\la12\ra}\int_{0\leq s\leq t\leq1}\dif s\,\dif t\,(1-t)\cos(t\,x\cdot k_1)\cos(s\,x\cdot k_2)\,. \ee
Finally we antisymmetrise in $1\leftrightarrow2$ as indicated in equation \eqref{eq:SDYM-perturbiner-dif}, to get
\be \label{eq:SDYM-perturbiner-final} \frac{\la1\beta\ra\la2\beta\ra[12]^2}{4\la12\ra}\int_0^1\dif s\,\int_0^1\dif t\,(1-\max(s,t))\cos(t\,x\cdot k_1)\cos(s\,x\cdot k_2)\,. \ee
This is the holomorphic collinear singularity in the self-dual Yang-Mills perturbiner on Eguchi-Hanson at first order in $c^2$, minus the term responsible for the shift in the zeroth order perturbiner to its curved counterpart. It appears in equation \eqref{eq:SDYM-perturbiner-singularity} of the bulk manuscript.


\subsection{Cancellation of logarithmic collinear singularities} \label{app:logarithmic-cancellation}

In this appendix we show that in the holomorphic collinear limit the subleading logarithmic singularities in $\cP_\mathrm{SDYM}^{(1)}(x;k_1,k_2)$ (as defined in equation \eqref{eq:SDYM-perturbiner-dif-1}) vanish. To this end, consider the contributions of the two terms in \eqref{eq:SDYM-vertex} to \eqref{eq:SDYM-singularity-integral} separately.

We already found that the first can be written as \eqref{eq:SDYM-simplified-term1}
\be \label{eq:SDYM-logarithmic-term1} \frac{1}{\pi^2\la\al1\ra}\int_0^1\frac{\dif s}{s^2}\,\la\al\p_{\lambda_2}\ra\la\beta\p_{\lambda_1}\ra\la\beta\p_{\lambda_2}\ra\big(\cI_2(x;k_-(s)) + \cI_2(x;k_+(s))\big)\,. \ee
A careful computation shows that the in the holomorphic collinear limit
\bea
&\la\al\p_{\lambda_2}\ra\la\beta\p_{\lambda_1}\ra\la\beta\p_{\lambda_2}\ra\cI_2(x;k_\pm(s))\sim \mathrm{simple~pole} \\
&+ \frac{s^2\pi^2[12]^2\log\la12\ra}{16}\int_0^1\dif t\,(1-t)^2\Big(2\la\beta1\ra\cos(t\,x\cdot k_\pm(s)) \\
& - 2[\ut1]\la\al1\ra\la\beta1\ra t\sin(t\,x\cdot k_\pm(s)) \mp 2\big([u2]\la\beta1\ra - [\ut2]\la\al1\ra\big)\la\beta2\ra st\sin(t\,x\cdot k_\pm(s)) \\
&\mp \la\al1\ra\la\beta1\ra\la \beta2\ra[12]x^2st(1-t)\cos(t\,x\cdot k_\pm(s))\Big) + \cO(1,\la12\ra\log\la12\ra)\,.
\eea
Therefore, the subleading logarithmic singularity in equation \eqref{eq:SDYM-logarithmic-term1} is
\bea \label{eq:SDYM-logarithmic-singularities1}
&\frac{[12]^2\log\la12\ra}{8\la\al1\ra}\int_0^1\dif s\,\int_0^1\dif t\,(1-t)^2\Big(2\la\beta1\ra\cos(t\,x\cdot k_1)\cos(st\,x\cdot k_2) \\
&- 2[\ut1]\la\al1\ra\la\beta1\ra t\sin(t\,x\cdot k_1)\cos(st\,x\cdot k_2) \\
&- 2\big([u2]\la\beta1\ra - [\ut2]\la\al1\ra\big)\la\beta2\ra st\cos(t\,x\cdot k_1)\sin(st\,x\cdot k_2) \\
&+ \la\al1\ra\la\beta1\ra\la\beta2\ra[12]x^2st(1-t)\sin(t\,x\cdot k_1)\sin(st\,x\cdot k_2)\Big)\,. 
\eea

The contribution of the second term can be written as \eqref{eq:SDYM-simplified-term2}
\be \label{eq:SDYM-logarithmic-term2} \frac{\la\al2\ra[12]}{4\pi^2\la\al1\ra}\int_0^1\frac{\dif s}{s}\la\beta\p_{\lambda_2}\ra^2\big(\cI_1(x;k_-(s))-\cI_1(x;k_+(s))\big)\,. \ee
In the holomorphic collinear limit
\bea
&\la\beta\p_{\lambda_2}\ra^2\cI_1(x,k_\pm(s)) \sim \mathrm{simple~pole} - \frac{s^2\pi^2\la\beta1\ra[12]\log\la12\ra}{4}\int_0^1\dif t\,t(1-t) \\
&\big(4[\ut2]\sin(t\,x\cdot k_\pm(s)) - \la\beta1\ra[12]x^2(1-t)\cos(t\,x\cdot k_\pm(s))\big) + \cO(1,\la12\ra\log\la12\ra)\,.
\eea
Hence, the subleading logarithmic singularity in equation \eqref{eq:SDYM-logarithmic-term2} is
\bea \label{eq:SDYM-logarithmic-singularities2} 
&- \frac{\la\al2\ra\la\beta1\ra[12]^2\log\la12\ra}{8\la\al1\ra}\int_0^1\dif s\,s\int_0^1\dif t\,t(1-t)\big(4[\ut2]\la\beta1\ra\cos(t\,x\cdot k_1)\sin(st\,x\cdot k_2) \\
&+ \la\beta1\ra[12]x^2(1-t)\sin(t\,x\cdot k_1)\sin(st\,x\cdot k_2)\big)\,. 
\eea

We can now proceed by combining like terms in equations \eqref{eq:SDYM-logarithmic-singularities1} and \eqref{eq:SDYM-logarithmic-singularities2}. For example, the terms involving $\sin(t\,x\cdot k_1)\sin(st\,x\cdot k_2)$ can be combined (invoking a Schouten and working modulo non-singular $\la12\ra\log\la12\ra$ terms) to give
\be
- \frac{\la\beta1\ra\la\beta2\ra x^2[12]^3\log\la12\ra}{8}\int_0^1\dif s\,s\int_0^1\dif t\,t^2(1-t)^2\sin(t\,x\cdot k_1)\sin(st\,x\cdot k_2) 
\ee
Exploiting a Schouten, the terms involving $\cos(t\,x\cdot k_1)\sin(st\,x\cdot k_2)$ sum to
\bea
&\frac{[12]^2\log\la12\ra}{4\la\al1\ra}\int_0^1\dif s\,s\int_0^1\dif t\,t(1-t)\cos(t\,x\cdot k_1)\sin(st\,x\cdot k_2) \\
&\qquad\qquad\qquad\big( - (x\cdot k_2)\la\beta1\ra + t([u2]\la\beta1\ra - [\ut2]\la\al1\ra)\la\beta2\ra\big)\,.
\eea
It's then natural to integrate by parts with respect to $s$ in the first of the above terms in order to eliminate $x\cdot k_2$. This gives
\bea
&\frac{\la \beta1\ra[12]^2\log\la12\ra}{4\la\al1\ra}\int_0^1\dif t\,(1-t)\cos(t\,x\cdot k_1)\cos(t\,x\cdot k_2) \\
&- \frac{\la\beta1\ra[12]^2\log\la12\ra}{4\la\al1\ra}\int_0^1\dif s\,\int_0^1\dif t\,(1-t)\cos(t\,x\cdot k_1)\cos(st\,x\cdot k_2)\,. \eea
Having performed these manipulations, and upon rescaling $s$ so that it now takes values in the range $[0,t]$, the total logarithmic singularity is
\bea
&\frac{[12]^2\log\la12\ra}{8\la\al1\ra}\int_{0\leq s\leq t\leq 1}\dif s\,\dif t\,(1-t)\big(- 2\la\beta1\ra\cos(t\,x\cdot k_1)\cos(s\,x\cdot k_2) \\
&\qquad\qquad\qquad\qquad- 2[\ut1]\la\al1\ra\la\beta1\ra(1-t)\sin(t\,x\cdot k_1)\cos(s\,x\cdot k_2) \\
&\qquad\qquad\qquad\qquad+ 2\big([u2]\la\beta1\ra - [\ut2]\la\al1\ra\big)\la\beta2\ra s\cos(t\,x\cdot k_1)\sin(s\,x\cdot k_2) \\
&\qquad\qquad\qquad\qquad- \la\al1\ra\la\beta1\ra\la\beta2\ra x^2[12]s(1-t)\sin(t\,x\cdot k_1)\sin(s\,x\cdot k_2)\big) \\
&+ \frac{\la \beta1\ra[12]^2\log\la12\ra}{4\la\al1\ra}\int_0^1\dif t\,(1-t)\cos(t\,x\cdot k_1)\cos(t\,x\cdot k_2)\,.
\eea
We can now iteratively integrate by parts with respect to $s,t$, so that the integrands become proportional to $\sin(t\,x\cdot k_1)\sin(st\,x\cdot k_2)$. As we do this we must take care to keep track of the boundary terms generated on the diagonal $s=t$. We have
\bea
&\int_{0\leq s\leq t\leq 1}\dif s\,\dif t\,(1-t)\cos(t\,x\cdot k_1)\cos(s\,x\cdot k_2) \\
&= - \frac{1}{2}(x\cdot k_1)(x\cdot k_2)\int_{0\leq s\leq t\leq1}\dif s\,\dif t\,s(1-t)^2\sin(t\,x\cdot k_1)\sin(t\,x\cdot k_2) \\
&+ \frac{1}{2}(x\cdot k_2)\int_0^1\dif t\,t(1-t)^2\cos(t\,x\cdot k_1)\sin(t\,x\cdot k_2) + \int_0^1\dif t\,t(1-t)\cos(t\,x\cdot k_1)\cos(t\,x\cdot k_2)\,, 
\eea
and similarly
\bea
&\int_{0\leq s\leq t\leq 1}\dif s\,\dif t\,(1-t)^2\sin(t\,x\cdot k_1)\cos(s\,x\cdot k_2) \\
&= (x\cdot k_2)\int_{0\leq s\leq t\leq 1}\dif s\,\dif t\,s(1-t)^2\sin(t\,x\cdot k_1)\sin(s\,x\cdot k_2) + \int_0^1\dif t\,t(1-t)^2\sin(t\,x\cdot k_1)\cos(t\,x\cdot k_2)\,,
\eea
as well as
\bea
&\int_{0\leq s\leq t\leq 1}\dif s\,\dif t\,s(1-t)\cos(t\,x\cdot k_1)\sin(s\,x\cdot k_2) \\
&= - \frac{1}{2}(x\cdot k_1)\int_{0\leq s\leq t\leq 1}\dif s\,\dif t\,s(1-t)^2\sin(t\,x\cdot k_1)\sin(s\,x\cdot k_2) + \frac{1}{2}\int_0^1\dif t\,t(1-t)^2\cos(t\,x\cdot k_1)\sin(t\,x\cdot k_2)\,.
\eea
The coefficient of the double integral
\be \int_{0\leq s\leq t\leq1}\dif s\,\dif t\,s(1-t)^2\sin(t\,x\cdot k_1)\sin(s\,x\cdot k_2) \ee
is, after some massaging,
\bea
&(x\cdot k_1)(x\cdot k_2)\la\beta1\ra - 2(x\cdot k_2)[\ut1]\la\al1\ra\la\beta1\ra - (x\cdot k_1)\big([u2]\la\beta1\ra - [\ut2]\la\al1\ra\big)\la\beta2\ra - \la\al1\ra\la\beta1\ra\la\beta2\ra[12]x^2 \\
&= (x\cdot k_1)((x\cdot k_2)\la\beta1\ra - [u2]\la\beta1\ra\la\beta2\ra + [\ut2]\la\al1\ra\la\beta2\ra) - 2(x\cdot k_2)[\ut1]\la\al1\ra\la\beta1\ra - \la\al1\ra\la\beta1\ra\la\beta2\ra[12]x^2 \\
&=  \big([u1]\la\beta1\ra - [\ut1]\la\al1\ra\big)[\ut2]\la12\ra\,,
\eea
so that it's contribution is non-singular. This leaves the boundary terms. The coefficient of
\be \int_0^1\dif t\,t(1-t)^2\cos(t\,x\cdot k_1)\sin(t\,x\cdot k_2) \ee
is proportional to
\bea
&(x\cdot k_2)\la\beta1\ra - [u2]\la\beta1\ra + [\ut2]\la\al1\ra\la\beta2\ra \\
&= [\ut2](\la\al1\ra\la\beta2\ra + \la\al2\ra\la\beta1\ra) = 2[\ut2]\la\al1\ra\la\beta2\ra + [\ut2]\la12\ra\,,
\eea
so that the full logarithmic singularity is
\bea
&\frac{[12]^2\log\la12\ra}{4\la\al1\ra\la\al2\ra}\int_0^1\dif t\,(1-t)^2\Bigg(\la\al2\ra\la\beta1\ra\cos(t\,x\cdot k_1)\cos(t\,x\cdot k_2)
\\
&\qquad\qquad- \la\al1\ra\la\al2\ra t\big([\ut1]\la\beta1\ra\sin(t\,x\cdot k_1)\cos(t\,x\cdot k_2) + (1\leftrightarrow2)\big)\Bigg)\,.
\eea
Upon antisymmetrising to get the perturbiner as indicated in equation \eqref{eq:SDYM-perturbiner-dif-1}, we find that above is non-singular.

\end{appendix}


\bibliographystyle{JHEP}
\bibliography{references}
	

\end{document}